\documentclass{aastex631}
\usepackage{graphics}
\usepackage{epstopdf}
\usepackage{float}
\usepackage{subfigure}
\usepackage{CJK}
\usepackage[T1]{fontenc}

\begin{document}
\begin{CJK*}{UTF8}{gbsn}
\setlength{\lineskiplimit}{0pt}
\setlength{\lineskip}{0pt}
\setlength{\abovedisplayskip}{6pt}  
\setlength{\belowdisplayskip}{6pt}
\setlength{\abovedisplayshortskip}{6pt}
\setlength{\belowdisplayshortskip}{6pt}

\title{Physical Parameters of 11,100 Short-Period ASAS-SN Eclipsing Contact Binaries} 

\correspondingauthor{Xu-Zhi Li(李旭志), Qing-Feng Zhu(朱青峰), Xu Ding(丁旭)}
\email{lixuzhi@ustc.edu.cn, zhuqf@ustc.edu.cn, dingxu@ynao.ac.cn}

\author[0000-0002-2783-2080]{Xu-Zhi Li(李旭志)}
\affiliation{CAS Key Laboratory for Research in Galaxies and Cosmology, Department of Astronomy, University of Science and Technology of China, Hefei, 230026, China}
\affiliation{School of Astronomy and Space Sciences, University of Science and Technology of China, Hefei, 230026, China}

\author[0000-0003-0694-8946]{Qing-Feng Zhu(朱青峰)}
\affiliation{CAS Key Laboratory for Research in Galaxies and Cosmology, Department of Astronomy, University of Science and Technology of China, Hefei, 230026, China}
\affiliation{School of Astronomy and Space Sciences, University of Science and Technology of China, Hefei, 230026, China}
\affiliation{Deep Space Exploration Laboratory, Hefei 230088, China}

\author[0000-0002-2427-161X]{Xu Ding(丁旭)}
\affiliation{Yunnan Observatories, Chinese Academy of Sciences, 396 Yangfangwang, Guandu District, Kunming, 650216, Yunnan, P. R. China}
\affiliation{Key Laboratory of the Structure and Evolution of Celestial Objects, Chinese Academy of Sciences, P. O. Box 110, 650216 Kunming, P. R. China}
\affiliation{Center for Astronomical Mega-Science, Chinese Academy of Sciences, 20A Datun Road, Chaoyang District, Beijing 100012, P. R. China}

\author{Xiao-Hui Xu(徐小慧)}
\affiliation{CAS Key Laboratory for Research in Galaxies and Cosmology, Department of Astronomy, University of Science and Technology of China, Hefei, 230026, China}
\affiliation{School of Astronomy and Space Sciences, University of Science and Technology of China, Hefei, 230026, China}

\author{Hang Zheng(郑航)}
\affiliation{CAS Key Laboratory for Research in Galaxies and Cosmology, Department of Astronomy, University of Science and Technology of China, Hefei, 230026, China}
\affiliation{School of Astronomy and Space Sciences, University of Science and Technology of China, Hefei, 230026, China}

\author{Jin-Sheng Qiu(邱锦盛)}
\affiliation{CAS Key Laboratory for Research in Galaxies and Cosmology, Department of Astronomy, University of Science and Technology of China, Hefei, 230026, China}
\affiliation{School of Astronomy and Space Sciences, University of Science and Technology of China, Hefei, 230026, China}

\author[0000-0002-8898-3894]{Ming-Chao Liu(刘明超)}
\affiliation{Yunnan Observatories, Chinese Academy of Sciences, 396 Yangfangwang, Guandu District, Kunming, 650216, Yunnan, P. R. China}
\affiliation{University of the Chinese Academy of Sciences, Yuquan Road 19, Shijingshan Block, Beijing, 100049, P. R. China}



\begin{abstract}

Starting from more than 11,200 short-period (less than 0.5 days) EW-type eclipsing binary candidates with the All-Sky Automated Survey for Supernovae (ASAS-SN) V-band light curves, we use MCMC and neural networks (NNs) to obtain the mass ratio ($q$), orbital inclination ($incl$), fill-out factor ($f$) and temperature ratio ($T_s/T_p$). After cross-matching with the Gaia DR3 database, the final sample contains parameters of 2,399 A-type and 8,712 W-type contact binaries (CBs). We present the distributions of parameters of these 11,111 short-period CBs. The mass ratio ($q$) and fill-out factor ($f$) are found to obey log-normal distributions, and the remaining parameters obey normal distributions. There is a significant period-temperature correlation of these CBs. Additionally, the temperature ratio (${T_s}$/${T_p}$) tends to increase as the orbital period decreases for W-type CBs. There is no significant correlation between them for A-type CBs. The mass ratio and fill-out factor ($q-f$) diagram suggest there is no significant correlation between these two parameters. A clear correlation exists between the mass ratio and radius ratio. The radius ratio increases with the mass ratio. Moreover, the deep fill-out CBs tend to fall on the upper boundary of the $q$$-$${R_s}$/${R_p}$ distribution, while the shallow fill-out CBs fall on the lower boundary.

\end{abstract}

\keywords{ Eclipsing binary stars(444) --- Contact binary stars(297) --- Fundamental properties of stars(555)}


\section{Introduction} \label{sec:intro}


Contact binary systems (CBs) are close binaries in which both components fill their Roche lobes and share a common envelope \citep{1941ApJ....93..133K,1959cbs..book.....K,1968ApJ...151.1123L,1968ApJ...153..877L}. Previous studies have revealed that CBs whose components have similar temperatures \citep{1941ApJ....93..133K} are caused by the exchange of energy and mass transfer through the common envelope \citep{2003ASPC..293...76W,2005ApJ...629.1055Y}.

Most of the eclipsing CBs are EW-type eclipsing binary systems according to the morphological classification of light curves \citep{1998stel.conf...81M,2020PASJ...72..103L}. Therefore, to understand the formation and evolution of CBs, EW-type eclipsing binaries are generally studied. The well-known period-luminosity-color (PLC) relations \citep{1967MmRAS..70..111E,1994PASP..106..462R,2016ApJ...832..138C,2018ApJS..237...28C,2017AJ....154..125M,2020MNRAS.493.4045J,2017RAA....17...87Q,2020RAA....20..163Q} of CBs are established by using different passband photometric data of EW-type eclipsing binaries. The PLC relations are different between early- and late-type CBs \citep{2020MNRAS.493.4045J}, and the PLC relations of different types of CBs have been extensively studied \citep{2011AcA....61..103G,2016AcA....66..421P,1994PASP..106..462R,2016ApJ...832..138C,2018ApJ...859..140C,2018ApJS..237...28C,2017AJ....154..125M,2020MNRAS.493.4045J,2017RAA....17...87Q,2020RAA....20..163Q}. 
In previous studies, early-type and late-type CBs are usually separated on the basis of their period.  Late-type CBs generally have a period of less than 0.5 days and are much more numerous than early-type CBs. 
\citet{2018ApJ...859..140C} defined the late-type CBs to have orbital periods $\log(P/d)\leqslant 0.25$, but \citet{2020MNRAS.493.4045J} think that a period separation of $\log(P/d)\leqslant 0.30$ is a better choice. In this paper, we focus on late-type CBs, which are usually short-period CBs.
Currently, statistical studies of CBs \citep{2017RAA....17...87Q,2019ApJS..244...43Z,2020ApJS..247...50S,2020MNRAS.493.4045J,2021AJ....161..176R,2021ApJS..254...10L,2022MNRAS.516.5003D} are becoming increasingly important thanks to the development of time-domain survey projects, e.g., the Catalina Sky Survey \citep [CSS;][]{2009ApJ...696..870D,2014ApJS..213....9D}, the Wide-field Infrared Survey Explorer \citep [WISE;][]{2010AJ....140.1868W,2018ApJS..237...28C}, the Northern Sky Variability Survey \citep [NSVS;][]{2004AJ....127.2436W,2006AJ....131..621G}, the Asteroid Terrestrial-impact Last Alert System \citep [ATLAS;][]{2018AJ....156..241H,2021AJ....161..176R}, the Optical Gravitational Lensing Experiment \citep [OGLE;][]{1992AcA....42..253U,1994AcA....44..317U,2016AcA....66..405S}, the Zwicky Transient Facility \citep [ZTF;][]{2019PASP..131f8003B,2020ApJS..249...18C}, the All-Sky Automated Survey for Supernovae \citep [ASAS-SN;][]{1997AcA....47..467P,2006MNRAS.368.1311P,2018MNRAS.477.3145J,2019MNRAS.486.1907J}, and some Space-based surveys like $Hipparcos$ \citep{1997yCat.1239....0E} , $Kepler$ \citep{2011AJ....141...83P,2014AJ....147...45C,2016AJ....151...68K}, $TESS$ \citep{2022ApJS..258...16P} and $Gaia$ \citep{2017A&A...606A..92M,2023A&A...674A..16M}.

The ASAS-SN \citep{2014ApJ...788...48S,
2018MNRAS.477.3145J,2019MNRAS.486.1907J,
2019MNRAS.485..961J,2020MNRAS.491...13J,
2021MNRAS.503..200J,2023MNRAS.519.5271C} 
is an automatic time-domain photometric sky survey project that currently consists of 24 telescopes distributed around the globe. More details can be seen on the websites \footnote{\url{https://www.astronomy.ohio-state.edu/asassn/index.shtml}}. The ASAS-SN photometric down to V $\sim$17 mag collected $\sim$2000 to over 7500 epochs of V-band observations per field. The ASAS-SN Variable Stars Database\footnote{\url{https://asas-sn.osu.edu/variables}}  has revealed $\sim$688,000 variable stars, $\sim$78,000 of which are classified EW-type eclipsing binaries.

In this paper, we use the ASAS-SN V-band photometric light curves of EW-type eclipsing binaries to study short-period CBs. In section 2, we describe the sample selection and light curve analysis methods. In section 3, we give the parameter distribution and some discussion of these CBs. Finally, the summary and conclusion are presented in section 4.

\section{Sample Selection and Analysis of the Light Curves} \label{sec:style}

\subsection{Sample Selection}

We used the ASAS-SN Variable Stars Database for target selection \citep{2014ApJ...788...48S,2018MNRAS.477.3145J,2019MNRAS.486.1907J,2019MNRAS.485..961J,2020MNRAS.491...13J,2021MNRAS.503..200J}. There are a total of 78,503 EW-type eclipsing binaries in the ASAS-SN Variable Stars Database. We search the short-period (less than 0.5 days) CB candidates under the following conditions. Based on the mean magnitude distribution of ASAS-SN \citep{2018MNRAS.477.3145J}, we select targets with mean V-band magnitude less than 15 mag, this includes the vast majority of targets. We also select targets with amplitudes greater than 0.2 mag because many ellipsoidal variable stars have amplitudes less than 0.2 mag and they are often mistaken for low-amplitude EW-type eclipsing binaries \citep{2013PASA...30...16D,2021NewA...8401539L,2022A&A...666A.142S}. The Laflfler-Kinman string length (LkSL) statistic \citep{1965ApJS...11..216L,2002A&A...386..763C} is a measure of scatter in the light curve of periodic variables. We used LKSL Statistic $\leqslant$ 0.1 to select the light curves of EW-type eclipsing binaries with good data quality. The class probabilities are given by \citet{2018MNRAS.477.3145J,2019MNRAS.486.1907J}, which are derived from a random forest classifier. They recommended that a class probability of $>$ 0.9 is the best classification. Under these conditions, we selected 11,282 candidate targets. We downloaded ASAS-SN V-band light curve data from the ASAS-SN Variable Stars Database for these 11,282 targets, of which 4 targets had no data, and 11,278 targets had valid data. 

\subsection{Neural Network and MCMC Algorithm Model}

Due to the development of time-domain photometric survey projects around the world, a large number of CB light curves have been released. It would be clearly impractical to use the Wilson-Devinney \citep{1971ApJ...166..605W,1990ApJ...356..613W,2007ApJ...661.1129V,2010ApJ...723.1469W,2012AJ....144...73W} or PHOEBE \citep{2016ApJS..227...29P} programs to derive the parameters of all these CBs. \citet{2008ApJ...687..542P} and \citet{2021PASJ...73..786D} used machine-learning methods to derive the parameters of detached binaries and contact binaries, respectively. These methods can speed up parameter derivation but cannot obtain the corresponding parameter errors. 

In this work, we use the neural network (NN) and the Markov chain Monte Carlo (MCMC) algorithm following the work in \citet{2021PASJ...73..786D,2022AJ....164..200D} to derive the parameters of contact binaries.
The NN model is trained with a set of light curves generated by Phoebe and known input parameters. Then, combined with the MCMC algorithm, the posterior distributions of the parameters can be quickly obtained.
We used Phoebe to generate two sets (with or without the third light) of synthetic light curves as samples. The third light is one value for each light curve, representing the effect of the third body on the system as an equivalent result. In this step, the filter band is set to $V$ because this work uses the ASAS-SN V-band data. The number of training light curves without and with the influence of the third light are 346,741 and 344,104, respectively. We select 10$\%$ of them as the test set. The model without the influence of the third light achieves a precision of 0.00045 magnitude in generating the light curves, while the model with the influence of the third light achieves a precision of 0.0002 magnitude in generating the light curves. Both models achieve a precision in generating the light curves that is less than 0.001 magnitude. Then, two models are obtained by training on the two sets of sample data. These two trained completed models can generate light curves according to the parameters of the CBs.

\subsection{Analysis of the Light Curves} 

We downloaded ASAS-SN V-band light curve data from the ASAS-SN Variable Stars Database. We also obtained the period and primary eclipse time $T_0$. Then, we phased these data and turned them into phase-magnitude data. To determine the fundamental photometric parameters of these ASAS-SN short-period CBs, the light curves were analyzed by using the NNs and MCMC algorithms.

We apply the two obtained models (with or without the third light) to obtain the parameters of the phased ASAS-SN light curves. Before using the models, we calculate the temperature for these CBs based on the given period-temperature relationship
by \citet{2020MNRAS.493.4045J} as follow:
\begin{equation}
    T_{eff}=6598+5260*\log_{10}(P/0.5) K.
\end{equation}
We use the parameter of goodness-of-fit ($R^2$) to evaluate the relevant results. The goodness-of-fit ($R^2$) is calculated based on the light curves generated by Phoebe and the original phased ASAS-SN light curves. If one target fits better under the third light model and the luminosity ratio of the third light 
\begin{equation}
    L_{3}/(L_{1}+L_{2}+L_{3})\geqslant 0.2,
\end{equation}
we consider the third light present and use the third light fitting model. The non-third-light model was used for the rest of the cases. 

We have listed four example targets, the observations and best-fitting solutions of the light curves are displayed in Figure 1. We select all targets with goodness-of-fit ($R^2$) greater than 0.8, with a total of 11,145 short-period ASAS-SN CBs. The sky distribution of the short-period CBs in ASAS-SN, colored by their period, is shown in Figure 2.

\begin{figure}[!htbp]
\centering  
\subfigure{
\label{Fig.sub.1}
\includegraphics[width=8cm]{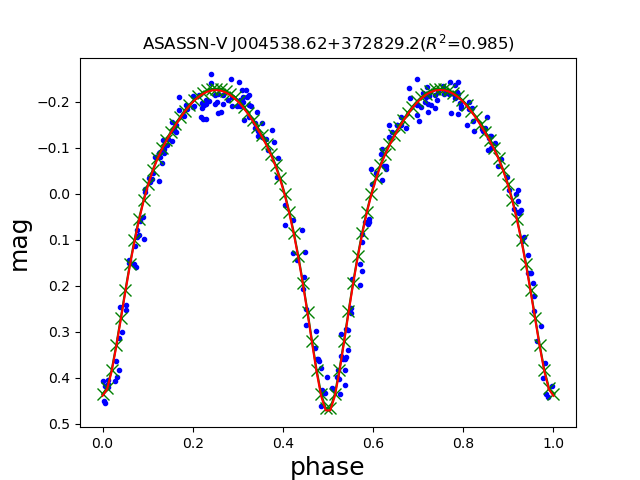}}
\subfigure{
\label{Fig.sub.2}
\includegraphics[width=8cm]{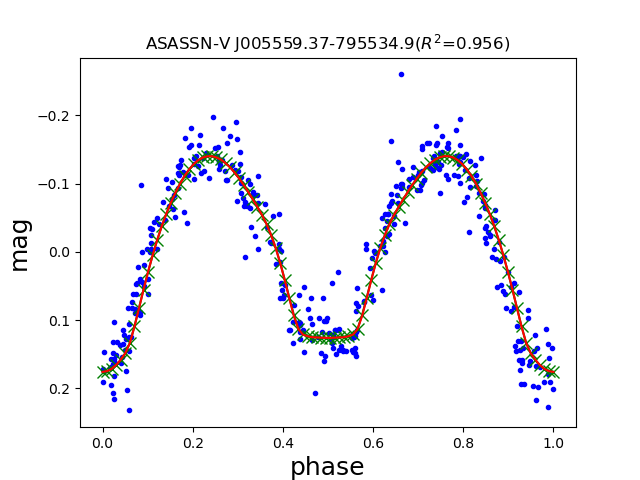}}

\subfigure{
\label{Fig.sub.3}
\includegraphics[width=8cm]{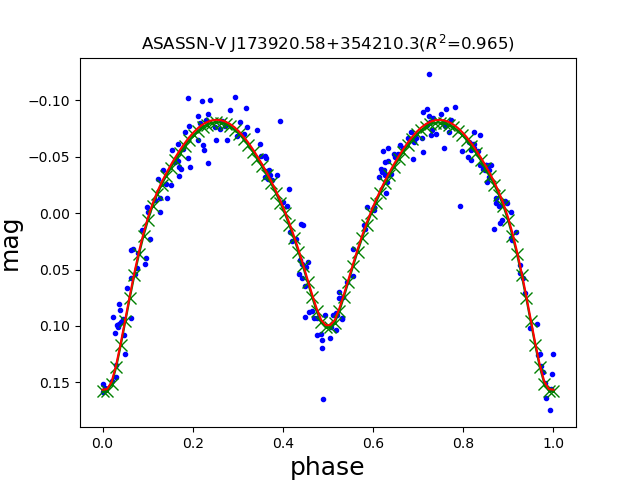}}
\subfigure{
\label{Fig.sub.4}
\includegraphics[width=8cm]{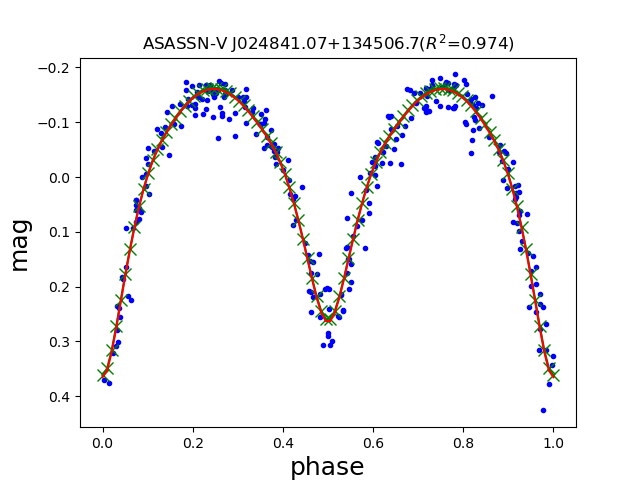}}
\caption{The blue dots represent the phased ASAS-SN light curves. Using the NNs and MCMC algorithm model to obtain the posterior distribution of the parameters as the input parameters. The red line represent the light curves generated by the NNs model with the input parameters. The green X line and symbol represent the light curves generated by Phoebe using the same input parameters. 
Top row: two example targets used the non-third-light model to get the parameters. Bottom row: two example targets used the third light model to get the parameters, and the luminosity ratio of the third light are about 68$\%$ and 27$\%$, respectively.}
\label{fig1}
\end{figure}

\begin{figure}[!htbp]
\plotone{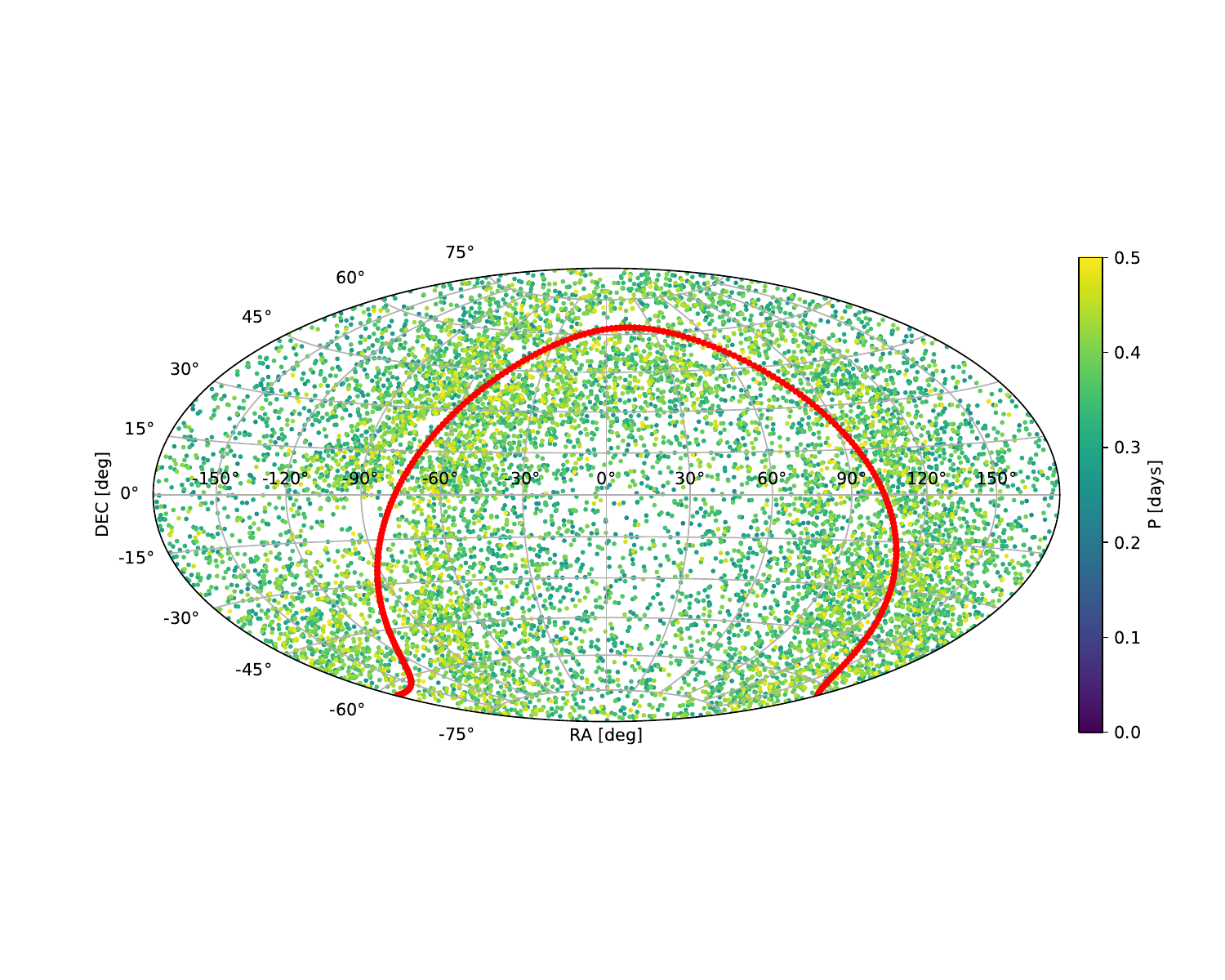}
\caption{Projected distribution of the 11,145 short-period ASAS-SN contact binaries in equatorial coordinates (Hammer projection). The points are colored by period. The red line indicates the Galactic plane.
\label{fig2}}
\end{figure}

\subsection{Final sample} 

Based on the analysis with the machine learning method (NN and MCMC algorithm model), we obtained the system parameters of ASAS-SN short-period CBs. The initial effective temperature of the first substar ($T_1$) is calculated by the period-temperature relationship \citep{2020MNRAS.493.4045J}. The orbital inclination ($incl$), mass ratio ($q$), temperature ratio ($T_2/T_1$), fill-out factor ($f$), third light ratio ($l_3$) and corresponding errors are derived by the machine learning method. The relative radii of the two substars ($r_1$, $r_2$) and luminosity ratio ($L_2/L_1$) are determined by Phoebe. 

We define mass ratio as values between 0 and 1, with the more massive star being the primary star and the less massive star being the secondary star. Thus, when the derived mass ratio ($q$) is greater than 1, we take the reciprocal as the final mass ratio. Similarly, the reciprocal of the temperature ratio $T_2/T_1$ and luminosity ratio $L_2/L_1$ should be $T_s/T_p$ and $L_s/L_p$, respectively. The radius $r_1$, $r_2$ should be exchanged as $R_s$, $R_p$, and the rest of the parameters do not need to be changed.

We cross-matched the CBs with Gaia DR3 \citep{2023A&A...674A...1G} using a matching radius of 5.0 arcsec to obtain the temperature of the primary stars ($T_p$). Finally, we matched a total of 11,111 targets, of which 9,621 could obtain the temperatures ($T_{eff}$) from Gaia DR3. For the remaining 1490 targets, we still use the calculated temperature by the period-temperature relationship \citep{2020MNRAS.493.4045J} as the primary star temperature. The final parameters for these CBs are given in the attached table in catalog format. The information listed in Table 1 is an introduction to the contents of catalog-1. 

\begin{table}[!htbp]\small
\begin{center}
\caption{Contents of the catalog-1 with 11,111 CBs}\label{Table2}
\begin{tabular}{cccl}\hline\hline
     Num   &  Label  &  Units  &  Explanations   \\
  \hline\noalign{\smallskip}
1    & name                   &           &      ASAS-SN Name  \\
2    & RA                     & $\circ$   &      RA (J2000)    \\
3    & Dec                    & $\circ$   &      DEC (J2000)   \\
4    & Period                 &  day      &      Orbital period  \\
5    & $T_p$                  &  $K$      &      Effective temperature of the primary star\\
6    & $incl$                 & $\circ$   &      Orbital inclination  \\
7    & $\sigma _{i}$          & $\circ$   &      Uncertainty in $incl$  \\
8    & $q$                    &           &      Mass ratio, $q=\frac{m_{s}}{m_{p}}\leqslant 1$  \\
9    & $\sigma _{q}$          &           &      Uncertainty in $q$  \\
10   & $f$                    &           &      Fill-out factor. $f =  \frac{\Omega_{in} - \Omega}{\Omega_{in} - \Omega_{out}}$,  \\
     & & & where $\Omega_{in}$ and $\Omega_{out}$ are the inner and outer Lagrangian surface potential values, \\
     & & & surface potential $\Omega$ $=$ $\Omega_{1}$ $=$ $\Omega_{2}$. \\
11    & $\sigma _{f}$          &           &      Uncertainty in $f$ \\
12   & $T_s/T_p$              &           &      Temperature ratio \\
13   & $T_s$                  & $K$       &      Effective temperature of the secondary star    \\
14   & Type                   &           &      CBs subtype classification (A or W)   \\
15   & $R_p$                  &           &      Relative radius of the primary star \\
16   & $R_s$                  &           &      Relative radius of the secondary star \\
17   & $L_s/L_p$              &           &      Luminosity ratio \\
18   & $l_3$                  &           &      Third light ratio $l_3=L_{3}/(L_{1}+L_{2}+L_{3})$ \\
19   & $R^2$                  &           &      Goodness of fit      \\
20   & DR3Name                &           &      Gaia DR3 ID  \\
21   & $T_{eff}$              & $K$       &     Effective temperature from Gaia DR3   \\
\hline
\end{tabular}
\end{center}
\end{table}
\clearpage

\section{Parameter distribution and Discussion} 

In the previous section, we obtained physical parameters of 11,111 ASAS-SN short-period CBs. The physical parameters include orbital period ($Period$), temperatures of the primary stars and the secondary stars ($T_p$, $T_s$), mass ratio ($q$), orbital inclination ($incl$), fill-out factor ($f$), temperature ratio ($T_s/T_p$), relative radii of the two substars ($R_p$, $R_s$), luminosity ratio ($L_s/L_p$). Here, we show the distribution of each parameter. 

\subsection{Parameter distribution} 
Figure 3 show the distribution of orbital periods ($p$), temperature ($T_p$, $T_s$), temperature ratio ($T_s/T_p$), orbital inclination ($incl$), fill-out factor ($f$), mass ratio ($q$) and relative radius ($R_p$, $R_s$) for the 11,111 ASAS-SN short-period CBs. The period distribution of ASAS-SN short-period CBs peaks at approximately 0.35 d, which is slightly shifted from the typical distribution of short-period EW-type eclipsing binaries with a peak at 0.31 d \citep{2020RAA....20..163Q,2020MNRAS.493.4045J}. This may be caused by the fact that some short-period ellipsoidal variable stars are usually classified as EW-type eclipsing binaries. Ellipsoidal variable stars and low amplitude EW-type eclipsing binaries are not easily distinguished directly by light curves \citep{2013PASA...30...16D,2021NewA...8401539L,2022A&A...666A.142S}. However, in this paper, the selection of our sample with amplitudes greater than 0.2 mag, which has led to the early exclusion of many ellipsoidal variable stars that have been misclassified as low amplitude EW-type eclipsing binaries. 

\begin{figure}[H]
\plotone{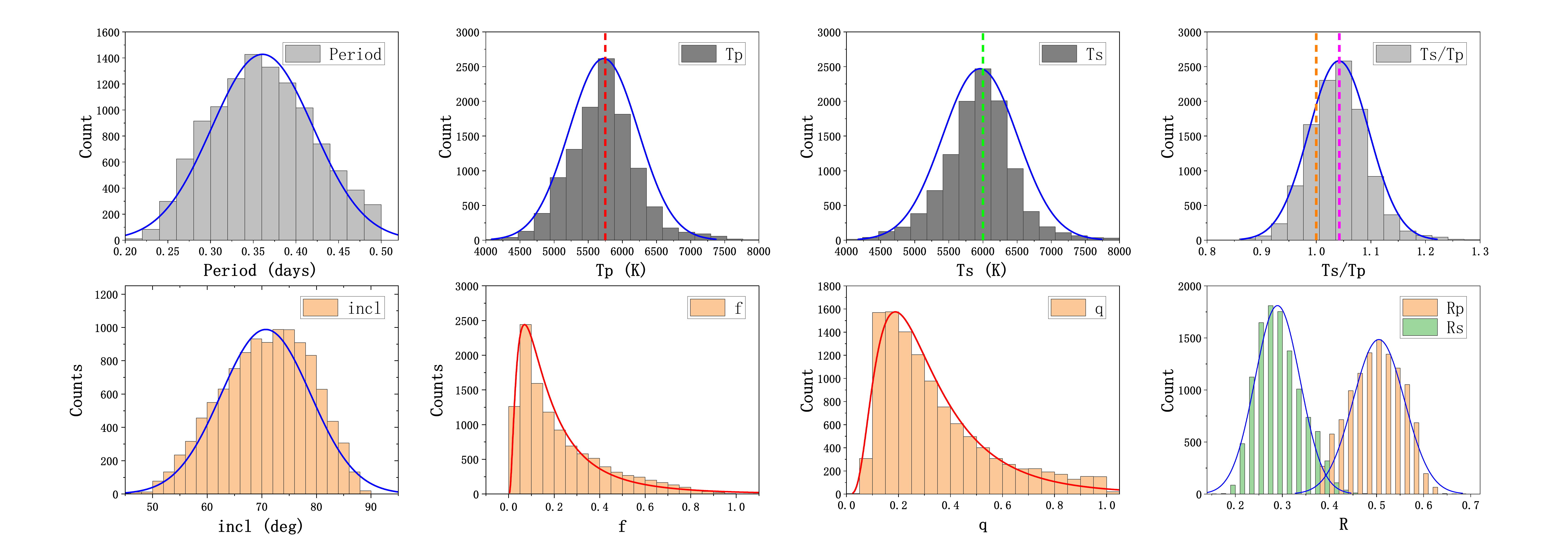}
\caption{Distribution of orbital periods (period), temperature ($T_p$, $T_s$), temperature ratio ($T_s/T_p$), orbital inclination ($incl$), fill-out factor ($f$), mass ratio ($q$) and relative radius ($R_p$, $R_s$) for the 11,111 ASAS-SN short-period CBs. The period distribution have a peak around 0.35 d. The temperature distribution of primary and secondary stars peaks at approximately 5750 $K$ (the red short dashed lines) and 6000 $K$ (the green short dashed lines), respectively. The orange short dashed lines represents a temperature ratio of 1, the left side represents the A-type CBs and the right side represents the W-type CBs. The distributions of the orbital period, temperature ($T_p$, $T_s$), temperature ratio ($T_s/T_p$), inclination ($incl$) and relative radius ($R_p$, $R_s$) roughly obey normal distributions (the blue solid line), the distributions of the mass ratio ($q$) and the fill-out factor ($f$) obey log-normal distributions (the red solid line).
\label{fig3}}
\end{figure}

The temperature distribution of ASAS-SN short-period CBs peaks at approximately 5750 $K$ (the red short dashed lines) for the primary star and 6000 $K$ (the green short dashed lines) for the secondary star. The peak of the secondary star temperature is higher than the peak of the primary star temperature, as can also be seen from the temperature ratio distribution. 
Traditionally, CBs have been classified into two subtypes based on temperature: A-type  and W-type \citep{1970VA.....12..217B}. In A-type the more massive star is hotter, while in W-type the more massive star is cooler. Our sample contains 2,399 A-type and 8,712 W-type CBs.

The orbital inclination distribution indicates that most CBs have orbital inclinations larger than $50^\circ$, which is likely due to selection effects, and CBs with high orbital inclination are more easily observed. Another reason is that our method is more reliable when used to fit targets with orbital inclinations greater than $50^\circ$ \citep{2021PASJ...73..786D}. The fill-out factor distribution of ASAS-SN short-period CBs indicates that CBs generally have shallow fill-out. The mass ratio distribution indicates that both extremely low mass ratio ($q \leqslant 0.1$) and large mass ratio ($q \geqslant 0.7$) CBs are less frequent. 

Very interestingly, according to the distribution of these short-period CB parameters, we find that only the distributions of the mass ratio and the fill-out factor obey log-normal distributions, while the rest of the parameters roughly obey normal distributions. We are confident that this is related to the evolution of contact binaries, but more simulation work is needed to confirm this.
 
\subsection{Period-Luminosity-Color Relations}

Figure 4 is a 2D kernel density graph. There is a clear correlation between the orbital period and the temperature of the primary star. The period-temperature relationship is actually the well-known period-color relationship or the so-called period-luminosity-color (PLC) relations \citep{1967MmRAS..70..111E,1994PASP..106..462R,2016ApJ...832..138C,2018ApJS..237...28C,2017AJ....154..125M,2020MNRAS.493.4045J,2017RAA....17...87Q,2020RAA....20..163Q}. Most CBs are located in a banded region rather than a simple linear PLC. This is related to the evolution of the common envelope of CBs, where the fill-out factor affects the PLC \citep{2020RAA....20..163Q}.

\begin{figure}[H]
\plotone{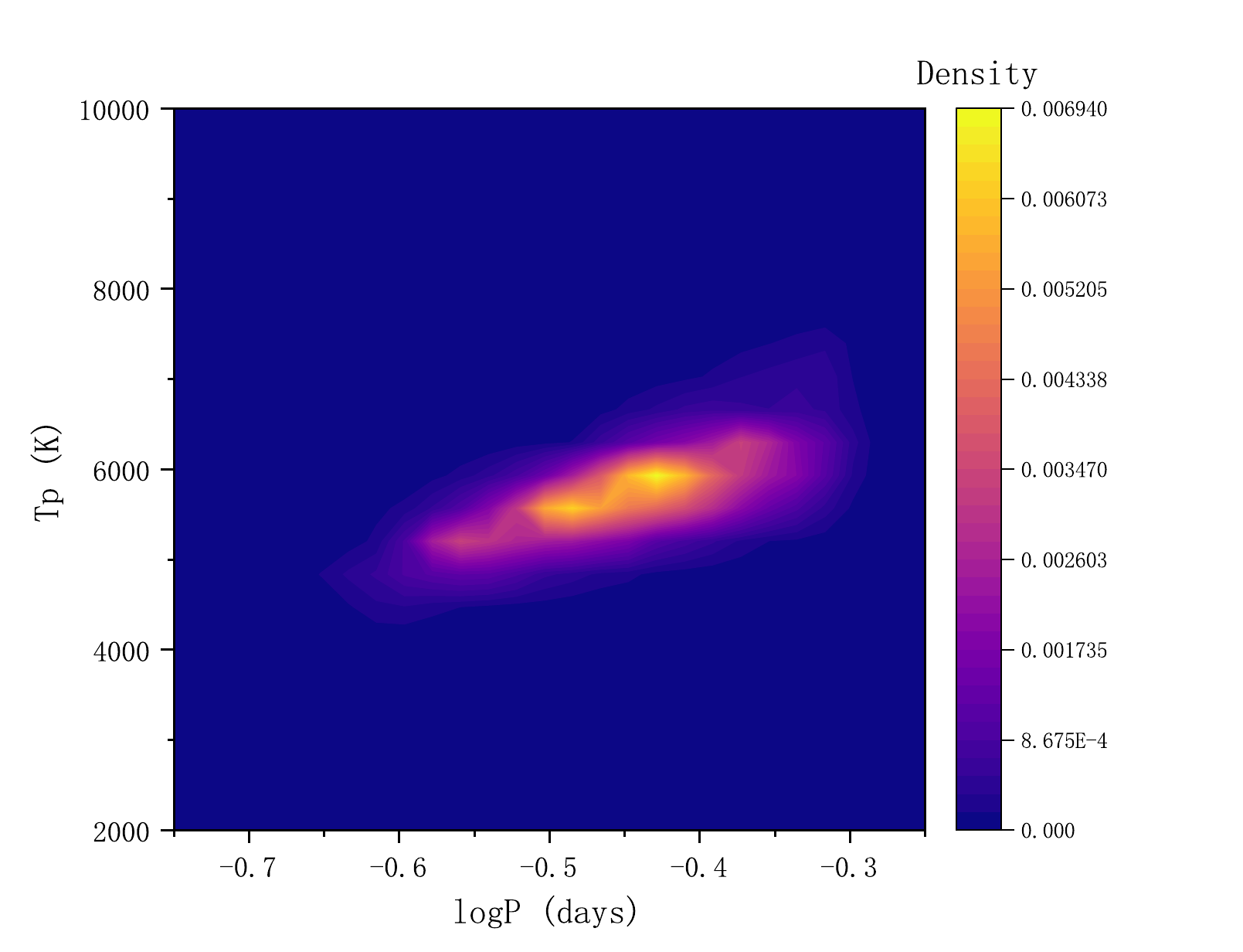}
\caption{The 2D kernel density graph between the orbital period and the temperature of the primary stars. It is show a clear correlation between the orbital period and the temperature of short-period CBs (usually are the late-type CBs), the longer the period, the higher the temperature.
\label{fig4}}
\end{figure}

\subsection{Period and Subtype CBs Evolution}

We plot the 2D kernel density graph between the orbital period and the temperature ratio, as shown in Figure 5. The period-temperature ratio seems to have some trend of correlation, as shown by the black line with arrows in Figure 5. As shown in Figure 5, there are far more W-type samples than A-type samples in the short-period CBs. For A-type CBs, there is no significant relationship between the temperature ratio and period, similar to a random distribution; however, for W-type CBs, the temperature ratio tends to increase as the orbital period decreases. This suggests that the evolution of A-type CBs is different from that of W-type CBs. Perhaps this could indicate that A-type CBs may be less evolved and suggest that W-type CBs evolved from A-type CBs \citep{2020PASJ...72..103L,2020MNRAS.492.4112Z}. Currently, the formation and evolution of CBs are still open issues \citep{2005ApJ...629.1055Y,2008MNRAS.390.1577G,2007ApJ...662..596L,2008MNRAS.387...97L,2009MNRAS.397..857S,2014MNRAS.437..185Y,2017RAA....17...87Q,2018ApJS..235....5Q,2020MNRAS.492.2731J,1996A&A...311..523M,2005JKAS...38...43A,2006MNRAS.373.1483E,2006MNRAS.370L..29G,2013MNRAS.430.2029Y}. 

\begin{figure}
\plotone{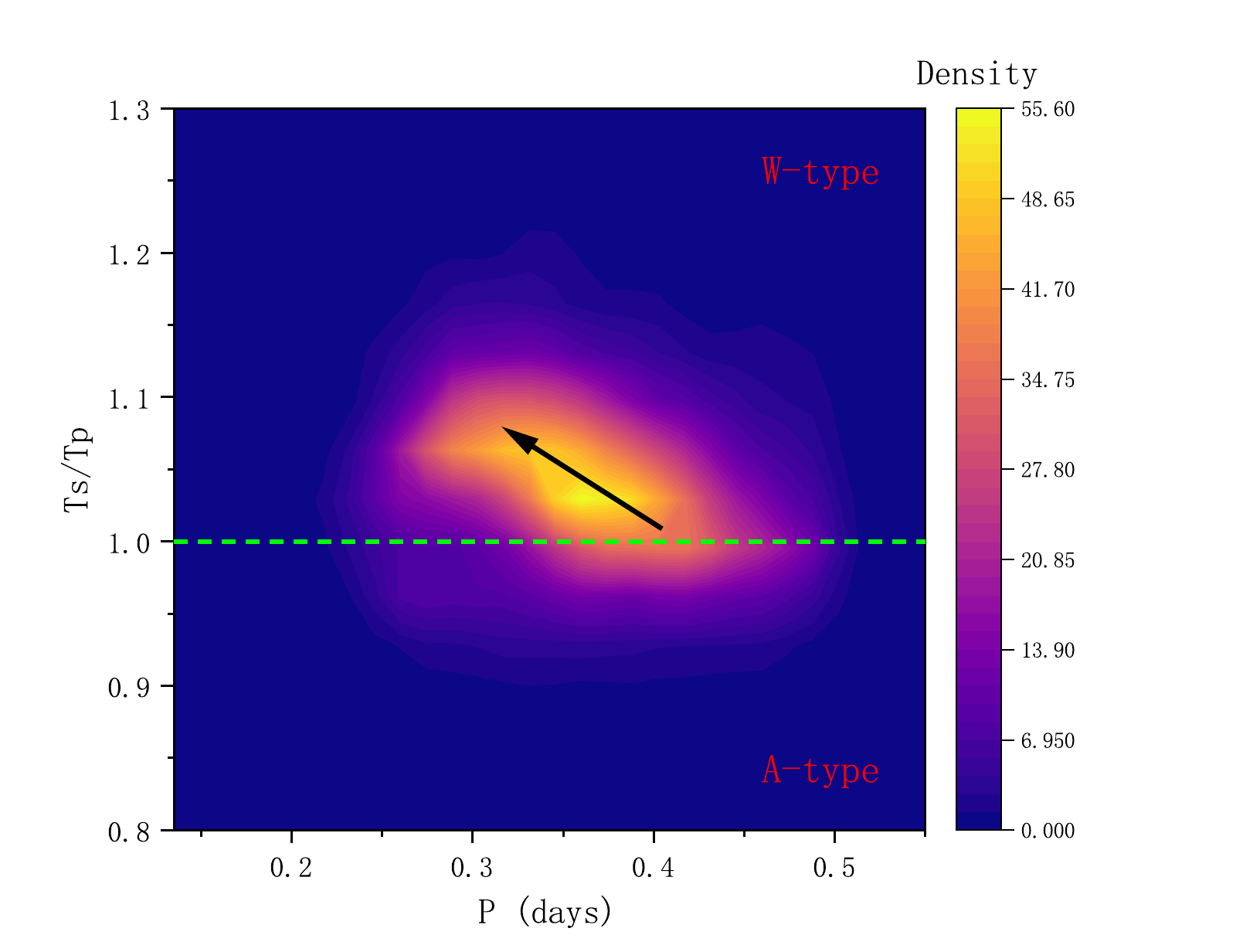}
\caption{The 2D kernel density graph between the orbital period and the temperature ratio. The green short dashed lines represent the temperature ratio as 1. The upper and lower parts of the short dashed lines represent the W-type and the A-type CBs, respectively.
\label{fig5}}
\end{figure}

Figure 6 shows the correlation between the orbital period and the temperature, mass ratio, orbital inclination and fill-out factor. This shows that they do not have a good correlation with the period, except for temperature. In addition, we find that the temperature, mass ratio, orbital inclination and fill-out factor are independently distributed from each other. There is no correlation among them.

\begin{figure}[H]
\plotone{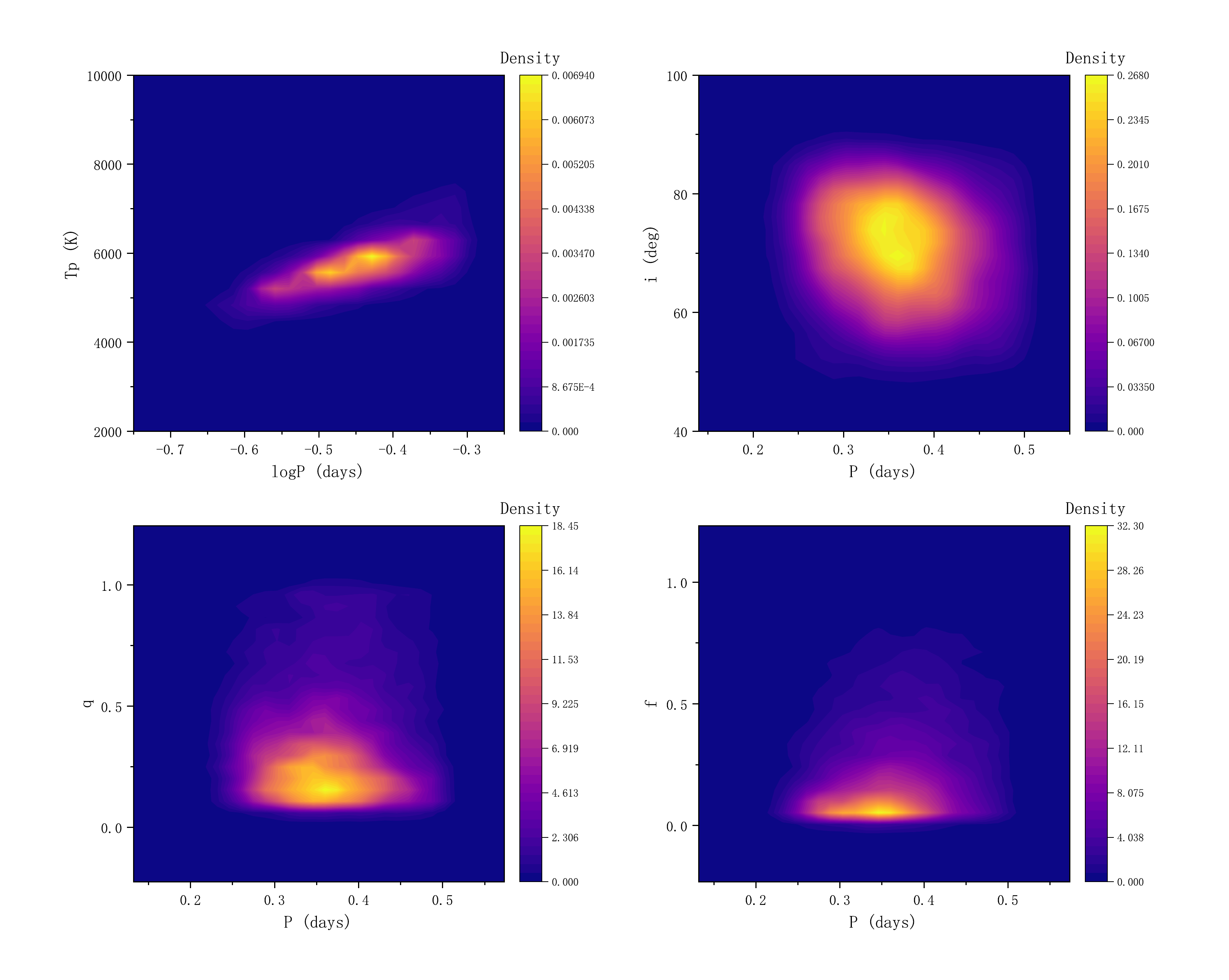}
\caption{The 2D kernel density graph between the orbital period and the temperature, mass ratio, orbital inclination and fill-out factor. There are no certain correlations between orbital period with mass ratio, orbital inclination and fill-out factor.
\label{fig6}}
\end{figure}

\subsection{q-f Diagram}

Figure 7 shows the correlation between the mass ratio and fill-out factor of ASAS-SN short-period CBs. As shown in Figure 3 and discussed in section 3.1, we find that the distributions of the mass ratio and the fill-out factor obey log-normal distributions. The correlation between the mass ratio and the fill-out factor is weak, as shown in Figure 7. However, we can also find much useful information in the $q-f$ diagram. 

In recent decades, there have been several studies on deep, low-mass ratio contact binaries (DLMRCBs) with mass ratios and fill-out factors of $q<0.25$ and $f>50\%$ \citep{2015AJ....150...69Y,2016ApJ...817..133Z,2021ApJ...922..122L,2022AJ....164..202L}, corresponding to the upper left region in Figure 7. Model calculations suggest that these DLMRCBs will finally merge into single stars, such as blue stragglers or FK-Com-type stars \citep{2006AcA....56..199S,2011AcA....61..139S}. V1309 Sco, the only observed sample, was a DLMRCB before the merge and was recently found as a blue straggler \citep{2011A&A...528A.114T,2016RAA....16...68Z,2019MNRAS.486.1220F}. Recently, \citet{2021MNRAS.502.2879G} and \citet{2022MNRAS.514.5528L} led a project focusing on DLMRCBs. They find that low mass-ratio systems seem more likely to eventually merge into single fast-roating stars. Our DLMRCB samples are located in the upper left region in Figure 7, and they should also be candidates for merging. 

As shown in Figure 7, there is no clear correlation between the mass ratio and fill-out factor. However, it show the upper limit of the fill-out factor increases as the mass ratio decreases, as shown by the green solid line in the figure, which is consistent with our previous results \citep{2020PASJ...72..103L}. 
In the previous study we had a sample of only 380 $Kepler$ CBs, this time we have a much larger sample.

We can also see in Figure 7 that there are very few samples in the upper right region that contain CBs with a deep fill-out factor and high mass ratio ($q>0.25$ and $f>50\%$). The common convective envelope-dominated mechanism \citep{2018MNRAS.474.5199L} can explain the presence of short period, deep fill-out factor and high mass ratio CBs. According to the CCE-dominated mechanism, $P$ will becomes very short when $f$ is large at a high value of $q$. 

\begin{figure}[H]
\plotone{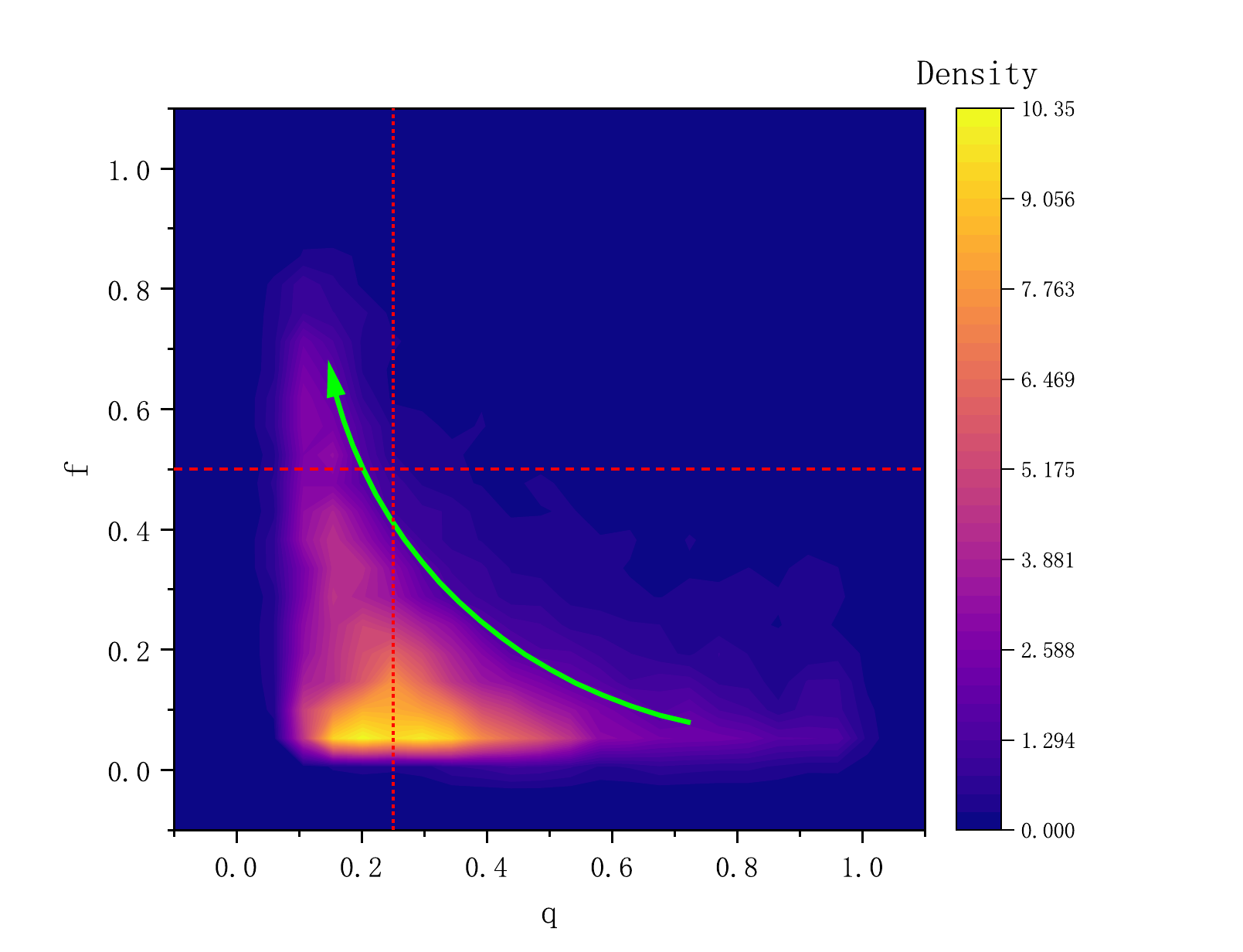}
\caption{The 2D kernel density graph between the mass ratio and fill-out factor of ASAS-SN short-period CBs. The red horizontal short dashed lines and vertical short dashed lines represent fill-out factor of 0.5 and mass ratio of 0.25, respectively. The green solid line indicates the trend between mass ratio ($q$) and fill-out factor ($f$). There are very few sample in the upper right region, thus CBs with deep fill-out factor, high mass ratio are more worthy of attention and study.
\label{fig7}}
\end{figure}

\subsection{Mass ratio - radius ratio Relationship}

The mass-radius relation ($M-R$) and mass-luminosity relation ($M-L$) of the primary and secondary stars of CBs have been studied by many authors \citep{2005ApJ...629.1055Y,2008MNRAS.387...97L,2013MNRAS.430.2029Y,2020PASJ...72..103L,2021ApJS..254...10L}. The $M-R$ relation of a CB system is significantly different from that of a single zero main sequence star. An approximate relation between the radii and masses of the components is $\frac{R_s}{R_p}=q^{0.46}$ \citep{1941ApJ....93..133K}, while for the zero main sequence star the relation is $\frac{R_s}{R_p}=q^{0.6}$ \citep{1968ApJ...151.1123L}.

We used the obtained data to fit the $q$ $-$ ${R_s}$/${R_p}$ relation, and the results are shown as the red solid line in Figure 8 as ${R_s}$/${R_p} $=$ q^{0.435}$, which is very close to the results in \citet{1941ApJ....93..133K}. The relationship between ${R_s}$/${R_p}$ and $q$ is monotonically increasing overall, but with dispersion at all $q$-value locations. We fit the upper and lower bounds of this region, as shown in the black and magenta solid lines in Figure 8. The black solid line has a relation of  ${R_s}$/${R_p} $=$ q^{0.38}$, while the magenta solid line has ${R_s}$/${R_p} $=$ q^{0.48}$. We find that the upper bound region is the deep fill-out CBs and the lower bound region is the shallow fill-out CBs. This suggests that the radius ratio is related to the fill-out factor of CBs for the same mass ratio \citep{2005ApJ...629.1055Y}.

\begin{figure}[H]
\plotone{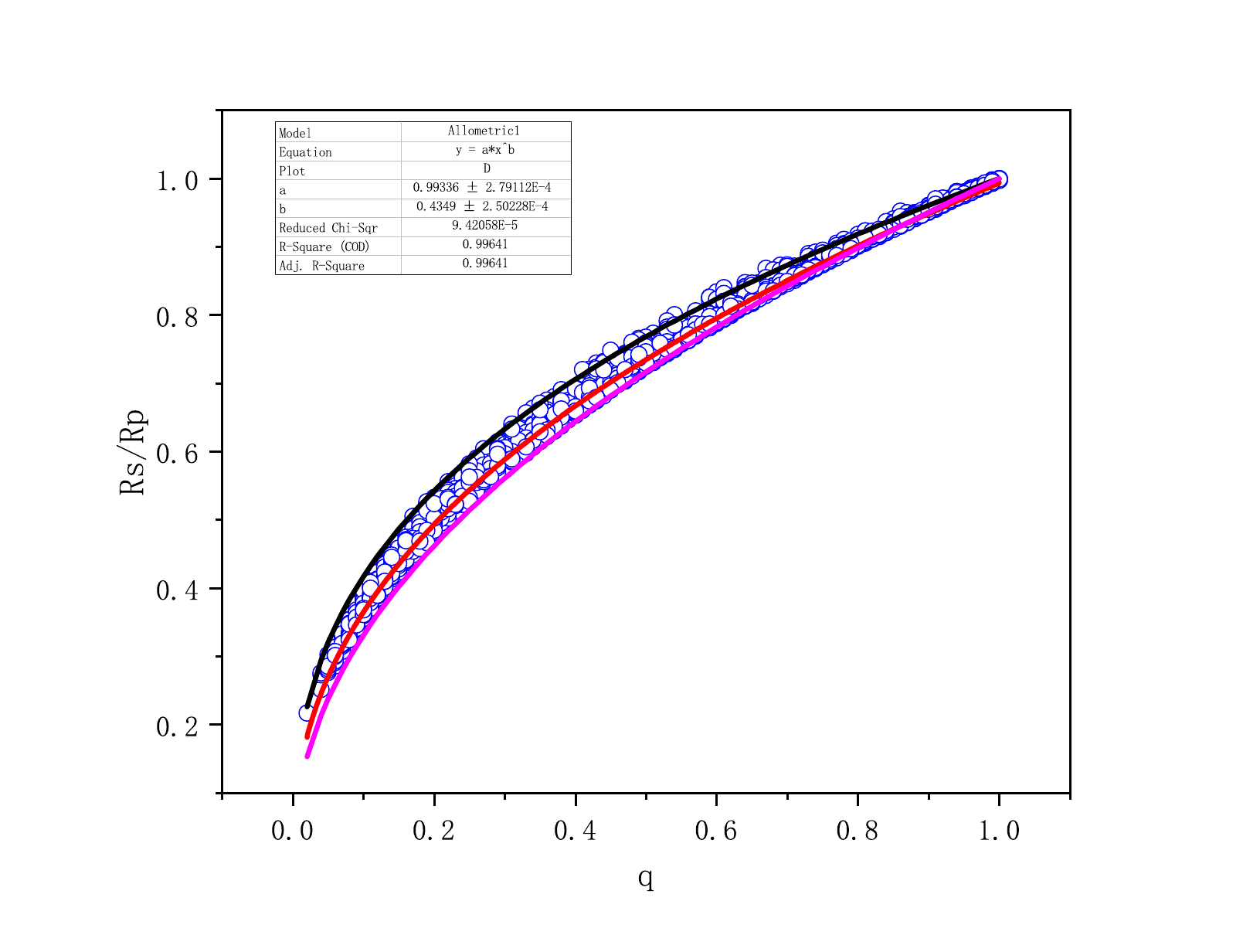}
\caption{Mass-radius Relationship of our short-period CBs. The red solid line represents the best fitting result as ${R_s}$/${R_p}$=$q^{0.435}$. The black solid line represents the fitting result as ${R_s}$/${R_p}$=$q^{0.38}$ for the upper bound, and the magenta solid line represents the fitting result as ${R_s}$/${R_p}$=$q^{0.48}$ for the lower bound. The upper bound region is the deep fill-out CBs and the lower bound region is the shallow fill-out CBs.
\label{fig8}}
\end{figure}

\section{Summary and Conclusions}

In this paper, we present the physical parameters of 11,111 short-period (see catalog-1) eclipsing contact binaries from the All-Sky Automated Survey for Supernovae (ASAS-SN) based on a machine learning method with neural networks (NNs) and Markov chain Monte Carlo (MCMC) algorithms. Based on the Gaia DR3 database, statistical work was performed for these ASAS-SN short-period CBs. Our main results and conclusions are summarized below.

1. After cross-matching with the Gaia DR3 database, our final sample contained 11,111 CBs, with a period less than 0.5 d, they were classified into two subtypes: 2,399 A-type and 8,712 W-type CBs.

2. We present the distributions of parameters of these short-period CBs, where the distributions of the mass ratio ($q$) and fill-out factor ($f$) are found to obey log-normal distributions, and the remaining parameters obey normal distributions.

3. There is a clear correlation between the orbital period and the temperature of short-period CBs, which is the well-known period-luminosity-color (PLC) relations.

4. Different subtypes of CBs seem to have different period and temperature ratio relationships. For A-type CBs, there is no significant correlation between the temperature ratio and period, similar to a random distribution; however, for W-type CBs, the temperature ratio tends to increase as the orbital period decreases. This likely suggests that the evolution of A-type CBs is different from that of W-type CBs.

5. The $q-f$ diagram indicated that there is no significant correlation between the mass ratio and fill-out factor. However, it can be seen that the fill-out factor increases as the mass ratio decreases. 

6. The mass ratio - radius ratio ($q$ $-$ ${R_s}$/${R_p}$) diagram indicated that there is a clear correlation between the mass ratio and radius ratio, the radius ratio increases with the mass ratio. And this correlation is also related to the fill-out factor of CBs, the deep fill-out CBs fall on the upper boundary of the $q$$-$${R_s}$/${R_p}$ distribution, while the shallow fill-out CBs fall on the lower boundary.

\begin{acknowledgments}

This work is supported by the National Key Research and Development Program of China (2023YFA1608100), the Anhui Provincial Natural Science Foundation (2308085QA35), the Strategic Priority Research Program of the Chinese Academy of Sciences (No. XDB 41000000), the China Postdoctoral Science Foundation (2021M703099) and the Fundamental Research Funds for the Central Universities.

This work made use of data from the ASAS-SN Variable Stars Database (\url{https://asas-sn.osu.edu/variables}). ASAS-SN is funded in part by the Gordon and Betty Moore Foundation through grants GBMF5490 and GBMF10501 to the Ohio State University and funded in part by the Alfred P. Sloan Foundation grant G-2021-14192. 

This work has made use of the cross-match service (\url{http://cdsxmatch.u-strasbg.fr/#tab=xmatch&}) provided by CDS, Strasbourg.

\end{acknowledgments}
%

\vspace{5mm}



\bibliography{ref}{}

\begin{thebibliography}{}
\expandafter\ifx\csname natexlab\endcsname\relax\def\natexlab#1{#1}\fi
\providecommand{\url}[1]{\href{#1}{#1}}
\providecommand{\dodoi}[1]{doi:~\href{http://doi.org/#1}{\nolinkurl{#1}}}
\providecommand{\doeprint}[1]{\href{http://ascl.net/#1}{\nolinkurl{http://ascl.net/#1}}}
\providecommand{\doarXiv}[1]{\href{https://arxiv.org/abs/#1}{\nolinkurl{https://arxiv.org/abs/#1}}}

\bibitem[{{Awadalla} \& {Hanna}(2005)}]{2005JKAS...38...43A}
{Awadalla}, N.~S., \& {Hanna}, M.~A. 2005, Journal of Korean Astronomical
  Society, 38, 43, \dodoi{10.5303/JKAS.2005.38.2.043}

\bibitem[{{Bellm} {et~al.}(2019){Bellm}, {Kulkarni}, {Barlow}, {Feindt},
  {Graham}, {Goobar}, {Kupfer}, {Ngeow}, {Nugent}, {Ofek}, {Prince}, {Riddle},
  {Walters}, \& {Ye}}]{2019PASP..131f8003B}
{Bellm}, E.~C., {Kulkarni}, S.~R., {Barlow}, T., {et~al.} 2019, \pasp, 131,
  068003, \dodoi{10.1088/1538-3873/ab0c2a}

\bibitem[{{Binnendijk}(1970)}]{1970VA.....12..217B}
{Binnendijk}, L. 1970, Vistas in Astronomy, 12, 217,
  \dodoi{10.1016/0083-6656(70)90041-3}

\bibitem[{{Chen} {et~al.}(2016){Chen}, {de Grijs}, \&
  {Deng}}]{2016ApJ...832..138C}
{Chen}, X., {de Grijs}, R., \& {Deng}, L. 2016, \apj, 832, 138,
  \dodoi{10.3847/0004-637X/832/2/138}

\bibitem[{{Chen} {et~al.}(2018{\natexlab{a}}){Chen}, {Deng}, {de Grijs},
  {Wang}, \& {Feng}}]{2018ApJ...859..140C}
{Chen}, X., {Deng}, L., {de Grijs}, R., {Wang}, S., \& {Feng}, Y.
  2018{\natexlab{a}}, \apj, 859, 140, \dodoi{10.3847/1538-4357/aabe83}

\bibitem[{{Chen} {et~al.}(2018{\natexlab{b}}){Chen}, {Wang}, {Deng}, {de
  Grijs}, \& {Yang}}]{2018ApJS..237...28C}
{Chen}, X., {Wang}, S., {Deng}, L., {de Grijs}, R., \& {Yang}, M.
  2018{\natexlab{b}}, \apjs, 237, 28, \dodoi{10.3847/1538-4365/aad32b}

\bibitem[{{Chen} {et~al.}(2020){Chen}, {Wang}, {Deng}, {de Grijs}, {Yang}, \&
  {Tian}}]{2020ApJS..249...18C}
{Chen}, X., {Wang}, S., {Deng}, L., {et~al.} 2020, \apjs, 249, 18,
  \dodoi{10.3847/1538-4365/ab9cae}

\bibitem[{{Christy} {et~al.}(2023){Christy}, {Jayasinghe}, {Stanek},
  {Kochanek}, {Thompson}, {Shappee}, {Holoien}, {Prieto}, {Dong}, \&
  {Giles}}]{2023MNRAS.519.5271C}
{Christy}, C.~T., {Jayasinghe}, T., {Stanek}, K.~Z., {et~al.} 2023, \mnras,
  519, 5271, \dodoi{10.1093/mnras/stac3801}

\bibitem[{{Clarke}(2002)}]{2002A&A...386..763C}
{Clarke}, D. 2002, \aap, 386, 763, \dodoi{10.1051/0004-6361:20020258}

\bibitem[{{Conroy} {et~al.}(2014){Conroy}, {Pr{\v{s}}a}, {Stassun}, {Orosz},
  {Fabrycky}, \& {Welsh}}]{2014AJ....147...45C}
{Conroy}, K.~E., {Pr{\v{s}}a}, A., {Stassun}, K.~G., {et~al.} 2014, \aj, 147,
  45, \dodoi{10.1088/0004-6256/147/2/45}

\bibitem[{{Dal} \& {Sipahi}(2013)}]{2013PASA...30...16D}
{Dal}, H.~A., \& {Sipahi}, E. 2013, \pasa, 30, e016,
  \dodoi{10.1017/pasa.2012.016}

\bibitem[{{Debski}(2022)}]{2022MNRAS.516.5003D}
{Debski}, B. 2022, \mnras, 516, 5003, \dodoi{10.1093/mnras/stac2190}

\bibitem[{{Ding} {et~al.}(2022){Ding}, {Ji}, {Li}, {Xiong}, {Cheng}, {Wang}, \&
  {Liu}}]{2022AJ....164..200D}
{Ding}, X., {Ji}, K., {Li}, X., {et~al.} 2022, \aj, 164, 200,
  \dodoi{10.3847/1538-3881/ac8e66}

\bibitem[{{Ding} {et~al.}(2021){Ding}, {Ji}, \& {Li}}]{2021PASJ...73..786D}
{Ding}, X., {Ji}, K.-F., \& {Li}, X.-Z. 2021, \pasj, 73, 786,
  \dodoi{10.1093/pasj/psab042}

\bibitem[{{Drake} {et~al.}(2009){Drake}, {Djorgovski}, {Mahabal}, {Beshore},
  {Larson}, {Graham}, {Williams}, {Christensen}, {Catelan}, {Boattini},
  {Gibbs}, {Hill}, \& {Kowalski}}]{2009ApJ...696..870D}
{Drake}, A.~J., {Djorgovski}, S.~G., {Mahabal}, A., {et~al.} 2009, \apj, 696,
  870, \dodoi{10.1088/0004-637X/696/1/870}

\bibitem[{{Drake} {et~al.}(2014){Drake}, {Graham}, {Djorgovski}, {Catelan},
  {Mahabal}, {Torrealba}, {Garc{\'\i}a-{\'A}lvarez}, {Donalek}, {Prieto},
  {Williams}, {Larson}, {Christen sen}, {Belokurov}, {Koposov}, {Beshore},
  {Boattini}, {Gibbs}, {Hill}, {Kowalski}, {Johnson}, \&
  {Shelly}}]{2014ApJS..213....9D}
{Drake}, A.~J., {Graham}, M.~J., {Djorgovski}, S.~G., {et~al.} 2014, \apjs,
  213, 9, \dodoi{10.1088/0067-0049/213/1/9}

\bibitem[{{Eggen}(1967)}]{1967MmRAS..70..111E}
{Eggen}, O.~J. 1967, \memras, 70, 111

\bibitem[{{Eker} {et~al.}(2006){Eker}, {Demircan}, {Bilir}, \&
  {Karata{\c{s}}}}]{2006MNRAS.373.1483E}
{Eker}, Z., {Demircan}, O., {Bilir}, S., \& {Karata{\c{s}}}, Y. 2006, \mnras,
  373, 1483, \dodoi{10.1111/j.1365-2966.2006.11073.x}

\bibitem[{{Esa}(1997)}]{1997yCat.1239....0E}
{Esa}, . 1997, VizieR Online Data Catalog, I/239

\bibitem[{{Ferreira} {et~al.}(2019){Ferreira}, {Saito}, {Minniti}, {Navarro},
  {Ramos}, {Smith}, \& {Lucas}}]{2019MNRAS.486.1220F}
{Ferreira}, T., {Saito}, R.~K., {Minniti}, D., {et~al.} 2019, \mnras, 486,
  1220, \dodoi{10.1093/mnras/stz878}

\bibitem[{{Gaia Collaboration} {et~al.}(2023){Gaia Collaboration}, {Vallenari},
  {Brown}, {Prusti}, {de Bruijne}, {Arenou}, {Babusiaux}, {Biermann},
  {Creevey}, {Ducourant}, {Evans}, {Eyer}, {Guerra}, {Hutton}, {Jordi},
  {Klioner}, {Lammers}, {Lindegren}, {Luri}, {Mignard}, {Panem}, {Pourbaix},
  {Randich}, {Sartoretti}, {Soubiran}, {Tanga}, {Walton}, {Bailer-Jones},
  {Bastian}, {Drimmel}, {Jansen}, {Katz}, {Lattanzi}, {van Leeuwen}, {Bakker},
  {Cacciari}, {Casta{\~n}eda}, {De Angeli}, {Fabricius}, {Fouesneau},
  {Fr{\'e}mat}, {Galluccio}, {Guerrier}, {Heiter}, {Masana}, {Messineo},
  {Mowlavi}, {Nicolas}, {Nienartowicz}, {Pailler}, {Panuzzo}, {Riclet}, {Roux},
  {Seabroke}, {Sordo}, {Th{\'e}venin}, {Gracia-Abril}, {Portell}, {Teyssier},
  {Altmann}, {Andrae}, {Audard}, {Bellas-Velidis}, {Benson}, {Berthier},
  {Blomme}, {Burgess}, {Busonero}, {Busso}, {C{\'a}novas}, {Carry}, {Cellino},
  {Cheek}, {Clementini}, {Damerdji}, {Davidson}, {de Teodoro}, {Nu{\~n}ez
  Campos}, {Delchambre}, {Dell'Oro}, {Esquej}, {Fern{\'a}ndez-Hern{\'a}ndez},
  {Fraile}, {Garabato}, {Garc{\'\i}a-Lario}, {Gosset}, {Haigron}, {Halbwachs},
  {Hambly}, {Harrison}, {Hern{\'a}ndez}, {Hestroffer}, {Hodgkin}, {Holl},
  {Jan{\ss}en}, {Jevardat de Fombelle}, {Jordan}, {Krone-Martins}, {Lanzafame},
  {L{\"o}ffler}, {Marchal}, {Marrese}, {Moitinho}, {Muinonen}, {Osborne},
  {Pancino}, {Pauwels}, {Recio-Blanco}, {Reyl{\'e}}, {Riello}, {Rimoldini},
  {Roegiers}, {Rybizki}, {Sarro}, {Siopis}, {Smith}, {Sozzetti}, {Utrilla},
  {van Leeuwen}, {Abbas}, {{\'A}brah{\'a}m}, {Abreu Aramburu}, {Aerts},
  {Aguado}, {Ajaj}, {Aldea-Montero}, {Altavilla}, {{\'A}lvarez}, {Alves},
  {Anders}, {Anderson}, {Anglada Varela}, {Antoja}, {Baines}, {Baker},
  {Balaguer-N{\'u}{\~n}ez}, {Balbinot}, {Balog}, {Barache}, {Barbato},
  {Barros}, {Barstow}, {Bartolom{\'e}}, {Bassilana}, {Bauchet}, {Becciani},
  {Bellazzini}, {Berihuete}, {Bernet}, {Bertone}, {Bianchi}, {Binnenfeld},
  {Blanco-Cuaresma}, {Blazere}, {Boch}, {Bombrun}, {Bossini}, {Bouquillon},
  {Bragaglia}, {Bramante}, {Breedt}, {Bressan}, {Brouillet}, {Brugaletta},
  {Bucciarelli}, {Burlacu}, {Butkevich}, {Buzzi}, {Caffau}, {Cancelliere},
  {Cantat-Gaudin}, {Carballo}, {Carlucci}, {Carnerero}, {Carrasco},
  {Casamiquela}, {Castellani}, {Castro-Ginard}, {Chaoul}, {Charlot}, {Chemin},
  {Chiaramida}, {Chiavassa}, {Chornay}, {Comoretto}, {Contursi}, {Cooper},
  {Cornez}, {Cowell}, {Crifo}, {Cropper}, {Crosta}, {Crowley}, {Dafonte},
  {Dapergolas}, {David}, {David}, {de Laverny}, {De Luise}, {De March}, {De
  Ridder}, {de Souza}, {de Torres}, {del Peloso}, {del Pozo}, {Delbo},
  {Delgado}, {Delisle}, {Demouchy}, {Dharmawardena}, {Di Matteo}, {Diakite},
  {Diener}, {Distefano}, {Dolding}, {Edvardsson}, {Enke}, {Fabre}, {Fabrizio},
  {Faigler}, {Fedorets}, {Fernique}, {Fienga}, {Figueras}, {Fournier},
  {Fouron}, {Fragkoudi}, {Gai}, {Garcia-Gutierrez}, {Garcia-Reinaldos},
  {Garc{\'\i}a-Torres}, {Garofalo}, {Gavel}, {Gavras}, {Gerlach}, {Geyer},
  {Giacobbe}, {Gilmore}, {Girona}, {Giuffrida}, {Gomel}, {Gomez},
  {Gonz{\'a}lez-N{\'u}{\~n}ez}, {Gonz{\'a}lez-Santamar{\'\i}a},
  {Gonz{\'a}lez-Vidal}, {Granvik}, {Guillout}, {Guiraud},
  {Guti{\'e}rrez-S{\'a}nchez}, {Guy}, {Hatzidimitriou}, {Hauser}, {Haywood},
  {Helmer}, {Helmi}, {Sarmiento}, {Hidalgo}, {Hilger}, {H{\l}adczuk}, {Hobbs},
  {Holland}, {Huckle}, {Jardine}, {Jasniewicz}, {Jean-Antoine Piccolo},
  {Jim{\'e}nez-Arranz}, {Jorissen}, {Juaristi Campillo}, {Julbe}, {Karbevska},
  {Kervella}, {Khanna}, {Kontizas}, {Kordopatis}, {Korn}, {K{\'o}sp{\'a}l},
  {Kostrzewa-Rutkowska}, {Kruszy{\'n}ska}, {Kun}, {Laizeau}, {Lambert},
  {Lanza}, {Lasne}, {Le Campion}, {Lebreton}, {Lebzelter}, {Leccia}, {Leclerc},
  {Lecoeur-Taibi}, {Liao}, {Licata}, {Lindstr{\o}m}, {Lister}, {Livanou},
  {Lobel}, {Lorca}, {Loup}, {Madrero Pardo}, {Magdaleno Romeo}, {Managau},
  {Mann}, {Manteiga}, {Marchant}, {Marconi}, {Marcos}, {Marcos Santos},
  {Mar{\'\i}n Pina}, {Marinoni}, {Marocco}, {Marshall}, {Martin Polo},
  {Mart{\'\i}n-Fleitas}, {Marton}, {Mary}, {Masip}, {Massari},
  {Mastrobuono-Battisti}, {Mazeh}, {McMillan}, {Messina}, {Michalik}, {Millar},
  {Mints}, {Molina}, {Molinaro}, {Moln{\'a}r}, {Monari}, {Mongui{\'o}},
  {Montegriffo}, {Montero}, {Mor}, {Mora}, {Morbidelli}, {Morel}, {Morris},
  {Muraveva}, {Murphy}, {Musella}, {Nagy}, {Noval}, {Oca{\~n}a}, {Ogden},
  {Ordenovic}, {Osinde}, {Pagani}, {Pagano}, {Palaversa}, {Palicio},
  {Pallas-Quintela}, {Panahi}, {Payne-Wardenaar}, {Pe{\~n}alosa Esteller},
  {Penttil{\"a}}, {Pichon}, {Piersimoni}, {Pineau}, {Plachy}, {Plum}, {Poggio},
  {Pr{\v{s}}a}, {Pulone}, {Racero}, {Ragaini}, {Rainer}, {Raiteri}, {Rambaux},
  {Ramos}, {Ramos-Lerate}, {Re Fiorentin}, {Regibo}, {Richards}, {Rios Diaz},
  {Ripepi}, {Riva}, {Rix}, {Rixon}, {Robichon}, {Robin}, {Robin}, {Roelens},
  {Rogues}, {Rohrbasser}, {Romero-G{\'o}mez}, {Rowell}, {Royer}, {Ruz Mieres},
  {Rybicki}, {Sadowski}, {S{\'a}ez N{\'u}{\~n}ez}, {Sagrist{\`a} Sell{\'e}s},
  {Sahlmann}, {Salguero}, {Samaras}, {Sanchez Gimenez}, {Sanna},
  {Santove{\~n}a}, {Sarasso}, {Schultheis}, {Sciacca}, {Segol}, {Segovia},
  {S{\'e}gransan}, {Semeux}, {Shahaf}, {Siddiqui}, {Siebert}, {Siltala},
  {Silvelo}, {Slezak}, {Slezak}, {Smart}, {Snaith}, {Solano}, {Solitro},
  {Souami}, {Souchay}, {Spagna}, {Spina}, {Spoto}, {Steele},
  {Steidelm{\"u}ller}, {Stephenson}, {S{\"u}veges}, {Surdej}, {Szabados},
  {Szegedi-Elek}, {Taris}, {Taylor}, {Teixeira}, {Tolomei}, {Tonello}, {Torra},
  {Torra}, {Torralba Elipe}, {Trabucchi}, {Tsounis}, {Turon}, {Ulla}, {Unger},
  {Vaillant}, {van Dillen}, {van Reeven}, {Vanel}, {Vecchiato}, {Viala},
  {Vicente}, {Voutsinas}, {Weiler}, {Wevers}, {Wyrzykowski}, {Yoldas}, {Yvard},
  {Zhao}, {Zorec}, {Zucker}, \& {Zwitter}}]{2023A&A...674A...1G}
{Gaia Collaboration}, {Vallenari}, A., {Brown}, A.~G.~A., {et~al.} 2023, \aap,
  674, A1, \dodoi{10.1051/0004-6361/202243940}

\bibitem[{{Gazeas} \& {St{\c{e}}pie{\'n}}(2008)}]{2008MNRAS.390.1577G}
{Gazeas}, K., \& {St{\c{e}}pie{\'n}}, K. 2008, \mnras, 390, 1577,
  \dodoi{10.1111/j.1365-2966.2008.13844.x}

\bibitem[{{Gazeas} \& {Niarchos}(2006)}]{2006MNRAS.370L..29G}
{Gazeas}, K.~D., \& {Niarchos}, P.~G. 2006, \mnras, 370, L29,
  \dodoi{10.1111/j.1745-3933.2006.00182.x}

\bibitem[{{Gazeas} {et~al.}(2021){Gazeas}, {Loukaidou}, {Niarchos},
  {Palafouta}, {Athanasopoulos}, {Liakos}, {Zola}, {Essam}, \&
  {Hakala}}]{2021MNRAS.502.2879G}
{Gazeas}, K.~D., {Loukaidou}, G.~A., {Niarchos}, P.~G., {et~al.} 2021, \mnras,
  502, 2879, \dodoi{10.1093/mnras/stab234}

\bibitem[{{Gettel} {et~al.}(2006){Gettel}, {Geske}, \&
  {McKay}}]{2006AJ....131..621G}
{Gettel}, S.~J., {Geske}, M.~T., \& {McKay}, T.~A. 2006, \aj, 131, 621,
  \dodoi{10.1086/498016}

\bibitem[{{Graczyk} {et~al.}(2011){Graczyk}, {Soszy{\'n}ski}, {Poleski},
  {Pietrzy{\'n}ski}, {Udalski}, {Szyma{\'n}ski}, {Kubiak}, {Wyrzykowski}, \&
  {Ulaczyk}}]{2011AcA....61..103G}
{Graczyk}, D., {Soszy{\'n}ski}, I., {Poleski}, R., {et~al.} 2011, \actaa, 61,
  103, \dodoi{10.48550/arXiv.1108.0446}

\bibitem[{{Heinze} {et~al.}(2018){Heinze}, {Tonry}, {Denneau}, {Flewelling},
  {Stalder}, {Rest}, {Smith}, {Smartt}, \& {Weiland}}]{2018AJ....156..241H}
{Heinze}, A.~N., {Tonry}, J.~L., {Denneau}, L., {et~al.} 2018, \aj, 156, 241,
  \dodoi{10.3847/1538-3881/aae47f}

\bibitem[{{Jayasinghe} {et~al.}(2018){Jayasinghe}, {Kochanek}, {Stanek},
  {Shappee}, {Holoien}, {Thompson}, {Prieto}, {Dong}, {Pawlak}, {Shields},
  {Pojmanski}, {Otero}, {Britt}, \& {Will}}]{2018MNRAS.477.3145J}
{Jayasinghe}, T., {Kochanek}, C.~S., {Stanek}, K.~Z., {et~al.} 2018, \mnras,
  477, 3145, \dodoi{10.1093/mnras/sty838}

\bibitem[{{Jayasinghe} {et~al.}(2019{\natexlab{a}}){Jayasinghe}, {Stanek},
  {Kochanek}, {Shappee}, {Holoien}, {Thompson}, {Prieto}, {Dong}, {Pawlak},
  {Pejcha}, {Shields}, {Pojmanski}, {Otero}, {Britt}, \&
  {Will}}]{2019MNRAS.486.1907J}
{Jayasinghe}, T., {Stanek}, K.~Z., {Kochanek}, C.~S., {et~al.}
  2019{\natexlab{a}}, \mnras, 486, 1907, \dodoi{10.1093/mnras/stz844}

\bibitem[{{Jayasinghe} {et~al.}(2019{\natexlab{b}}){Jayasinghe}, {Stanek},
  {Kochanek}, {Shappee}, {Holoien}, {Thompson}, {Prieto}, {Dong}, {Pawlak},
  {Pejcha}, {Shields}, {Pojmanski}, {Otero}, {Hurst}, {Britt}, \&
  {Will}}]{2019MNRAS.485..961J}
---. 2019{\natexlab{b}}, \mnras, 485, 961, \dodoi{10.1093/mnras/stz444}

\bibitem[{{Jayasinghe} {et~al.}(2020{\natexlab{a}}){Jayasinghe}, {Stanek},
  {Kochanek}, {Shappee}, {Pinsonneault}, {Holoien}, {Thompson}, {Prieto},
  {Pawlak}, {Pejcha}, {Pojmanski}, {Otero}, {Hurst}, \&
  {Will}}]{2020MNRAS.493.4045J}
---. 2020{\natexlab{a}}, \mnras, 493, 4045, \dodoi{10.1093/mnras/staa518}

\bibitem[{{Jayasinghe} {et~al.}(2020{\natexlab{b}}){Jayasinghe}, {Stanek},
  {Kochanek}, {Shappee}, {Holoien}, {Thompson}, {Prieto}, {Dong}, {Pawlak},
  {Pejcha}, {Shields}, {Pojmanski}, {Otero}, {Hurst}, {Britt}, \&
  {Will}}]{2020MNRAS.491...13J}
---. 2020{\natexlab{b}}, \mnras, 491, 13, \dodoi{10.1093/mnras/stz2711}

\bibitem[{{Jayasinghe} {et~al.}(2021){Jayasinghe}, {Kochanek}, {Stanek},
  {Shappee}, {Holoien}, {Thompson}, {Prieto}, {Dong}, {Pawlak}, {Pejcha},
  {Pojmanski}, {Otero}, {Hurst}, \& {Will}}]{2021MNRAS.503..200J}
{Jayasinghe}, T., {Kochanek}, C.~S., {Stanek}, K.~Z., {et~al.} 2021, \mnras,
  503, 200, \dodoi{10.1093/mnras/stab114}

\bibitem[{{Jiang}(2020)}]{2020MNRAS.492.2731J}
{Jiang}, D. 2020, \mnras, 492, 2731, \dodoi{10.1093/mnras/stz3578}

\bibitem[{{Kirk} {et~al.}(2016){Kirk}, {Conroy}, {Pr{\v{s}}a}, {Abdul-Masih},
  {Kochoska}, {Matijevi{\v{c}}}, {Hambleton}, {Barclay}, {Bloemen}, {Boyajian},
  {Doyle}, {Fulton}, {Hoekstra}, {Jek}, {Kane}, {Kostov}, {Latham}, {Mazeh},
  {Orosz}, {Pepper}, {Quarles}, {Ragozzine}, {Shporer}, {Southworth},
  {Stassun}, {Thompson}, {Welsh}, {Agol}, {Derekas}, {Devor}, {Fischer},
  {Green}, {Gropp}, {Jacobs}, {Johnston}, {LaCourse}, {Saetre}, {Schwengeler},
  {Toczyski}, {Werner}, {Garrett}, {Gore}, {Martinez}, {Spitzer}, {Stevick},
  {Thomadis}, {Vrijmoet}, {Yenawine}, {Batalha}, \&
  {Borucki}}]{2016AJ....151...68K}
{Kirk}, B., {Conroy}, K., {Pr{\v{s}}a}, A., {et~al.} 2016, \aj, 151, 68,
  \dodoi{10.3847/0004-6256/151/3/68}

\bibitem[{{Kopal}(1959)}]{1959cbs..book.....K}
{Kopal}, Z. 1959, {Close binary systems}

\bibitem[{{Kuiper}(1941)}]{1941ApJ....93..133K}
{Kuiper}, G.~P. 1941, \apj, 93, 133, \dodoi{10.1086/144252}

\bibitem[{{Lafler} \& {Kinman}(1965)}]{1965ApJS...11..216L}
{Lafler}, J., \& {Kinman}, T.~D. 1965, \apjs, 11, 216, \dodoi{10.1086/190116}

\bibitem[{{Latkovi{\'c}} {et~al.}(2021){Latkovi{\'c}}, {{\v{C}}eki}, \&
  {Lazarevi{\'c}}}]{2021ApJS..254...10L}
{Latkovi{\'c}}, O., {{\v{C}}eki}, A., \& {Lazarevi{\'c}}, S. 2021, \apjs, 254,
  10, \dodoi{10.3847/1538-4365/abeb23}

\bibitem[{{Li} {et~al.}(2022){Li}, {Gao}, {Liu}, {Gao}, {Li}, {Chen}, \&
  {Sun}}]{2022AJ....164..202L}
{Li}, K., {Gao}, X., {Liu}, X.-Y., {et~al.} 2022, \aj, 164, 202,
  \dodoi{10.3847/1538-3881/ac8ff2}

\bibitem[{{Li} {et~al.}(2021){Li}, {Xia}, {Kim}, {Hu}, {Guo}, {Jeong}, {Chen},
  \& {Gao}}]{2021ApJ...922..122L}
{Li}, K., {Xia}, Q.-Q., {Kim}, C.-H., {et~al.} 2021, \apj, 922, 122,
  \dodoi{10.3847/1538-4357/ac242f}

\bibitem[{{Li} {et~al.}(2007){Li}, {Zhang}, {Han}, \&
  {Jiang}}]{2007ApJ...662..596L}
{Li}, L., {Zhang}, F., {Han}, Z., \& {Jiang}, D. 2007, \apj, 662, 596,
  \dodoi{10.1086/517909}

\bibitem[{{Li} {et~al.}(2008){Li}, {Zhang}, {Han}, {Jiang}, \&
  {Jiang}}]{2008MNRAS.387...97L}
{Li}, L., {Zhang}, F., {Han}, Z., {Jiang}, D., \& {Jiang}, T. 2008, \mnras,
  387, 97, \dodoi{10.1111/j.1365-2966.2008.12736.x}

\bibitem[{{Li} \& {Liu}(2021)}]{2021NewA...8401539L}
{Li}, X.-Z., \& {Liu}, L. 2021, \na, 84, 101539,
  \dodoi{10.1016/j.newast.2020.101539}

\bibitem[{{Li} {et~al.}(2020){Li}, {Liu}, \& {Zhu}}]{2020PASJ...72..103L}
{Li}, X.-Z., {Liu}, L., \& {Zhu}, L.-Y. 2020, \pasj, 72, 103,
  \dodoi{10.1093/pasj/psaa104}

\bibitem[{{Liu} {et~al.}(2018){Liu}, {Qian}, \& {Xiong}}]{2018MNRAS.474.5199L}
{Liu}, L., {Qian}, S.~B., \& {Xiong}, X. 2018, \mnras, 474, 5199,
  \dodoi{10.1093/mnras/stx3138}

\bibitem[{{Loukaidou} {et~al.}(2022){Loukaidou}, {Gazeas}, {Palafouta},
  {Athanasopoulos}, {Zola}, {Liakos}, {Niarchos}, {Hakala}, {Essam}, \&
  {Hatzidimitriou}}]{2022MNRAS.514.5528L}
{Loukaidou}, G.~A., {Gazeas}, K.~D., {Palafouta}, S., {et~al.} 2022, \mnras,
  514, 5528, \dodoi{10.1093/mnras/stab3424}

\bibitem[{{Lucy}(1968{\natexlab{a}})}]{1968ApJ...151.1123L}
{Lucy}, L.~B. 1968{\natexlab{a}}, \apj, 151, 1123, \dodoi{10.1086/149510}

\bibitem[{{Lucy}(1968{\natexlab{b}})}]{1968ApJ...153..877L}
---. 1968{\natexlab{b}}, \apj, 153, 877, \dodoi{10.1086/149712}

\bibitem[{{Maceroni} \& {van't Veer}(1996)}]{1996A&A...311..523M}
{Maceroni}, C., \& {van't Veer}, F. 1996, \aap, 311, 523

\bibitem[{{Mateo} \& {Rucinski}(2017)}]{2017AJ....154..125M}
{Mateo}, N.~M., \& {Rucinski}, S.~M. 2017, \aj, 154, 125,
  \dodoi{10.3847/1538-3881/aa8453}

\bibitem[{{Mol{\'\i}k}(1998)}]{1998stel.conf...81M}
{Mol{\'\i}k}, P. 1998, in 20th Stellar Conference of the Czech and Slovak
  Astronomical Institutes, ed. J.~{Dusek}, 81

\bibitem[{{Mowlavi} {et~al.}(2017){Mowlavi}, {Lecoeur-Ta{\"\i}bi}, {Holl},
  {Rimoldini}, {Barblan}, {Pr{\v{s}}a}, {Kochoska}, {S{\"u}veges}, {Eyer},
  {Nienartowicz}, {Jevardat}, {Charnas}, {Guy}, \&
  {Audard}}]{2017A&A...606A..92M}
{Mowlavi}, N., {Lecoeur-Ta{\"\i}bi}, I., {Holl}, B., {et~al.} 2017, \aap, 606,
  A92, \dodoi{10.1051/0004-6361/201730613}

\bibitem[{{Mowlavi} {et~al.}(2023){Mowlavi}, {Holl}, {Lecoeur-Ta{\"\i}bi},
  {Barblan}, {Kochoska}, {Pr{\v{s}}a}, {Mazeh}, {Rimoldini}, {Gavras},
  {Audard}, {Jevardat de Fombelle}, {Nienartowicz}, {Garc{\'\i}a-Lario}, \&
  {Eyer}}]{2023A&A...674A..16M}
{Mowlavi}, N., {Holl}, B., {Lecoeur-Ta{\"\i}bi}, I., {et~al.} 2023, \aap, 674,
  A16, \dodoi{10.1051/0004-6361/202245330}

\bibitem[{{Paczy{\'n}ski} {et~al.}(2006){Paczy{\'n}ski}, {Szczygie{\l}},
  {Pilecki}, \& {Pojma{\'n}ski}}]{2006MNRAS.368.1311P}
{Paczy{\'n}ski}, B., {Szczygie{\l}}, D.~M., {Pilecki}, B., \& {Pojma{\'n}ski},
  G. 2006, \mnras, 368, 1311, \dodoi{10.1111/j.1365-2966.2006.10223.x}

\bibitem[{{Pawlak} {et~al.}(2016){Pawlak}, {Soszy{\'n}ski}, {Udalski},
  {Szyma{\'n}ski}, {Wyrzykowski}, {Ulaczyk}, {Poleski}, {Pietrukowicz},
  {Koz{\l}owski}, {Skowron}, {Skowron}, {Mr{\'o}z}, \&
  {Hamanowicz}}]{2016AcA....66..421P}
{Pawlak}, M., {Soszy{\'n}ski}, I., {Udalski}, A., {et~al.} 2016, \actaa, 66,
  421, \dodoi{10.48550/arXiv.1612.06394}

\bibitem[{{Pojmanski}(1997)}]{1997AcA....47..467P}
{Pojmanski}, G. 1997, \actaa, 47, 467.
\newblock \doarXiv{astro-ph/9712146}

\bibitem[{{Pr{\v{s}}a} {et~al.}(2008){Pr{\v{s}}a}, {Guinan}, {Devinney},
  {DeGeorge}, {Bradstreet}, {Giammarco}, {Alcock}, \&
  {Engle}}]{2008ApJ...687..542P}
{Pr{\v{s}}a}, A., {Guinan}, E.~F., {Devinney}, E.~J., {et~al.} 2008, \apj, 687,
  542, \dodoi{10.1086/591783}

\bibitem[{{Pr{\v{s}}a} {et~al.}(2011){Pr{\v{s}}a}, {Batalha}, {Slawson},
  {Doyle}, {Welsh}, {Orosz}, {Seager}, {Rucker}, {Mjaseth}, {Engle}, {Conroy},
  {Jenkins}, {Caldwell}, {Koch}, \& {Borucki}}]{2011AJ....141...83P}
{Pr{\v{s}}a}, A., {Batalha}, N., {Slawson}, R.~W., {et~al.} 2011, \aj, 141, 83,
  \dodoi{10.1088/0004-6256/141/3/83}

\bibitem[{{Pr{\v{s}}a} {et~al.}(2016){Pr{\v{s}}a}, {Conroy}, {Horvat}, {Pablo},
  {Kochoska}, {Bloemen}, {Giammarco}, {Hambleton}, \&
  {Degroote}}]{2016ApJS..227...29P}
{Pr{\v{s}}a}, A., {Conroy}, K.~E., {Horvat}, M., {et~al.} 2016, \apjs, 227, 29,
  \dodoi{10.3847/1538-4365/227/2/29}

\bibitem[{{Pr{\v{s}}a} {et~al.}(2022){Pr{\v{s}}a}, {Kochoska}, {Conroy},
  {Eisner}, {Hey}, {IJspeert}, {Kruse}, {Fleming}, {Johnston}, {Kristiansen},
  {LaCourse}, {Mortensen}, {Pepper}, {Stassun}, {Torres}, {Abdul-Masih},
  {Chakraborty}, {Gagliano}, {Guo}, {Hambleton}, {Hong}, {Jacobs}, {Jones},
  {Kostov}, {Lee}, {Omohundro}, {Orosz}, {Page}, {Powell}, {Rappaport}, {Reed},
  {Schnittman}, {Schwengeler}, {Shporer}, {Terentev}, {Vanderburg}, {Welsh},
  {Caldwell}, {Doty}, {Jenkins}, {Latham}, {Ricker}, {Seager}, {Schlieder},
  {Shiao}, {Vanderspek}, \& {Winn}}]{2022ApJS..258...16P}
{Pr{\v{s}}a}, A., {Kochoska}, A., {Conroy}, K.~E., {et~al.} 2022, \apjs, 258,
  16, \dodoi{10.3847/1538-4365/ac324a}

\bibitem[{{Qian} {et~al.}(2017){Qian}, {He}, {Zhang}, {Zhu}, {Shi}, {Zhao}, \&
  {Zhou}}]{2017RAA....17...87Q}
{Qian}, S.-B., {He}, J.-J., {Zhang}, J., {et~al.} 2017, Research in Astronomy
  and Astrophysics, 17, 087, \dodoi{10.1088/1674-4527/17/8/87}

\bibitem[{{Qian} {et~al.}(2018){Qian}, {Zhang}, {He}, {Zhu}, {Zhao}, {Shi},
  {Zhou}, \& {Han}}]{2018ApJS..235....5Q}
{Qian}, S.~B., {Zhang}, J., {He}, J.~J., {et~al.} 2018, \apjs, 235, 5,
  \dodoi{10.3847/1538-4365/aaa601}

\bibitem[{{Qian} {et~al.}(2020){Qian}, {Zhu}, {Liu}, {Zhang}, {Shi}, {He}, \&
  {Zhang}}]{2020RAA....20..163Q}
{Qian}, S.-B., {Zhu}, L.-Y., {Liu}, L., {et~al.} 2020, Research in Astronomy
  and Astrophysics, 20, 163, \dodoi{10.1088/1674-4527/20/10/163}

\bibitem[{{Ren} {et~al.}(2021){Ren}, {de Grijs}, {Zhang}, {Deng}, {Chen},
  {Matsunaga}, {Liu}, {Sun}, {Maehara}, {Ukita}, \&
  {Kobayashi}}]{2021AJ....161..176R}
{Ren}, F., {de Grijs}, R., {Zhang}, H., {et~al.} 2021, \aj, 161, 176,
  \dodoi{10.3847/1538-3881/abe30e}

\bibitem[{{Rucinski}(1994)}]{1994PASP..106..462R}
{Rucinski}, S.~M. 1994, \pasp, 106, 462, \dodoi{10.1086/133401}

\bibitem[{{Shappee} {et~al.}(2014){Shappee}, {Prieto}, {Grupe}, {Kochanek},
  {Stanek}, {De Rosa}, {Mathur}, {Zu}, {Peterson}, {Pogge}, {Komossa}, {Im},
  {Jencson}, {Holoien}, {Basu}, {Beacom}, {Szczygie{\l}}, {Brimacombe},
  {Adams}, {Campillay}, {Choi}, {Contreras}, {Dietrich}, {Dubberley},
  {Elphick}, {Foale}, {Giustini}, {Gonzalez}, {Hawkins}, {Howell}, {Hsiao},
  {Koss}, {Leighly}, {Morrell}, {Mudd}, {Mullins}, {Nugent}, {Parrent},
  {Phillips}, {Pojmanski}, {Rosing}, {Ross}, {Sand}, {Terndrup}, {Valenti},
  {Walker}, \& {Yoon}}]{2014ApJ...788...48S}
{Shappee}, B.~J., {Prieto}, J.~L., {Grupe}, D., {et~al.} 2014, \apj, 788, 48,
  \dodoi{10.1088/0004-637X/788/1/48}

\bibitem[{{Skarka} {et~al.}(2022){Skarka}, {{\v{Z}}{\'a}k}, {Fedurco},
  {Paunzen}, {Henzl}, {Ma{\v{s}}ek}, {Karjalainen}, {Sanchez Arias},
  {S{\'o}dor}, {Auer}, {Kab{\'a}th}, {Karjalainen}, {Li{\v{s}}ka}, \&
  {{\v{S}}tegner}}]{2022A&A...666A.142S}
{Skarka}, M., {{\v{Z}}{\'a}k}, J., {Fedurco}, M., {et~al.} 2022, \aap, 666,
  A142, \dodoi{10.1051/0004-6361/202244037}

\bibitem[{{Soszy{\'n}ski} {et~al.}(2016){Soszy{\'n}ski}, {Pawlak},
  {Pietrukowicz}, {Udalski}, {Szyma{\'n}ski}, {Wyrzykowski}, {Ulaczyk},
  {Poleski}, {Koz{\l}owski}, {Skowron}, {Skowron}, {Mr{\'o}z}, \&
  {Hamanowicz}}]{2016AcA....66..405S}
{Soszy{\'n}ski}, I., {Pawlak}, M., {Pietrukowicz}, P., {et~al.} 2016, \actaa,
  66, 405.
\newblock \doarXiv{1701.03105}

\bibitem[{{St{\c{e}}pie{\'n}}(2009)}]{2009MNRAS.397..857S}
{St{\c{e}}pie{\'n}}, K. 2009, \mnras, 397, 857,
  \dodoi{10.1111/j.1365-2966.2009.14981.x}

\bibitem[{{Stepien}(2006)}]{2006AcA....56..199S}
{Stepien}, K. 2006, \actaa, 56, 199.
\newblock \doarXiv{astro-ph/0510464}

\bibitem[{{St{\k{e}}pie{\'n}}(2011)}]{2011AcA....61..139S}
{St{\k{e}}pie{\'n}}, K. 2011, \actaa, 61, 139.
\newblock \doarXiv{1105.2645}

\bibitem[{{Sun} {et~al.}(2020){Sun}, {Chen}, {Deng}, \& {de
  Grijs}}]{2020ApJS..247...50S}
{Sun}, W., {Chen}, X., {Deng}, L., \& {de Grijs}, R. 2020, \apjs, 247, 50,
  \dodoi{10.3847/1538-4365/ab7894}

\bibitem[{{Tylenda} {et~al.}(2011){Tylenda}, {Hajduk}, {Kami{\'n}ski},
  {Udalski}, {Soszy{\'n}ski}, {Szyma{\'n}ski}, {Kubiak}, {Pietrzy{\'n}ski},
  {Poleski}, {Wyrzykowski}, \& {Ulaczyk}}]{2011A&A...528A.114T}
{Tylenda}, R., {Hajduk}, M., {Kami{\'n}ski}, T., {et~al.} 2011, \aap, 528,
  A114, \dodoi{10.1051/0004-6361/201016221}

\bibitem[{{Udalski} {et~al.}(1994){Udalski}, {Kubiak}, {Szymanski}, {Kaluzny},
  {Mateo}, \& {Krzeminski}}]{1994AcA....44..317U}
{Udalski}, A., {Kubiak}, M., {Szymanski}, M., {et~al.} 1994, \actaa, 44, 317

\bibitem[{{Udalski} {et~al.}(1992){Udalski}, {Szymanski}, {Kaluzny}, {Kubiak},
  \& {Mateo}}]{1992AcA....42..253U}
{Udalski}, A., {Szymanski}, M., {Kaluzny}, J., {Kubiak}, M., \& {Mateo}, M.
  1992, \actaa, 42, 253

\bibitem[{{Van Hamme} \& {Wilson}(2007)}]{2007ApJ...661.1129V}
{Van Hamme}, W., \& {Wilson}, R.~E. 2007, \apj, 661, 1129,
  \dodoi{10.1086/517870}

\bibitem[{{Webbink}(2003)}]{2003ASPC..293...76W}
{Webbink}, R.~F. 2003, in Astronomical Society of the Pacific Conference
  Series, Vol. 293, 3D Stellar Evolution, ed. S.~{Turcotte}, S.~C. {Keller}, \&
  R.~M. {Cavallo}, 76.
\newblock \doarXiv{astro-ph/0304420}

\bibitem[{{Wilson}(1990)}]{1990ApJ...356..613W}
{Wilson}, R.~E. 1990, \apj, 356, 613, \dodoi{10.1086/168867}

\bibitem[{{Wilson}(2012)}]{2012AJ....144...73W}
---. 2012, \aj, 144, 73, \dodoi{10.1088/0004-6256/144/3/73}

\bibitem[{{Wilson} \& {Devinney}(1971)}]{1971ApJ...166..605W}
{Wilson}, R.~E., \& {Devinney}, E.~J. 1971, \apj, 166, 605,
  \dodoi{10.1086/150986}

\bibitem[{{Wilson} {et~al.}(2010){Wilson}, {Van Hamme}, \&
  {Terrell}}]{2010ApJ...723.1469W}
{Wilson}, R.~E., {Van Hamme}, W., \& {Terrell}, D. 2010, \apj, 723, 1469,
  \dodoi{10.1088/0004-637X/723/2/1469}

\bibitem[{{Wo{\'z}niak} {et~al.}(2004){Wo{\'z}niak}, {Vestrand}, {Akerlof},
  {Balsano}, {Bloch}, {Casperson}, {Fletcher}, {Gisler}, {Kehoe}, {Kinemuchi},
  {Lee}, {Marshall}, {McGowan}, {McKay}, {Rykoff}, {Smith}, {Szymanski}, \&
  {Wren}}]{2004AJ....127.2436W}
{Wo{\'z}niak}, P.~R., {Vestrand}, W.~T., {Akerlof}, C.~W., {et~al.} 2004, \aj,
  127, 2436, \dodoi{10.1086/382719}

\bibitem[{{Wright} {et~al.}(2010){Wright}, {Eisenhardt}, {Mainzer}, {Ressler},
  {Cutri}, {Jarrett}, {Kirkpatrick}, {Padgett}, {McMillan}, {Skrutskie},
  {Stanford}, {Cohen}, {Walker}, {Mather}, {Leisawitz}, {Gautier}, {McLean},
  {Benford}, {Lonsdale}, {Blain}, {Mendez}, {Irace}, {Duval}, {Liu}, {Royer},
  {Heinrichsen}, {Howard}, {Shannon}, {Kendall}, {Walsh}, {Larsen}, {Cardon},
  {Schick}, {Schwalm}, {Abid}, {Fabinsky}, {Naes}, \&
  {Tsai}}]{2010AJ....140.1868W}
{Wright}, E.~L., {Eisenhardt}, P. R.~M., {Mainzer}, A.~K., {et~al.} 2010, \aj,
  140, 1868, \dodoi{10.1088/0004-6256/140/6/1868}

\bibitem[{{Yakut} \& {Eggleton}(2005)}]{2005ApJ...629.1055Y}
{Yakut}, K., \& {Eggleton}, P.~P. 2005, \apj, 629, 1055, \dodoi{10.1086/431300}

\bibitem[{{Yang} \& {Qian}(2015)}]{2015AJ....150...69Y}
{Yang}, Y.-G., \& {Qian}, S.-B. 2015, \aj, 150, 69,
  \dodoi{10.1088/0004-6256/150/3/69}

\bibitem[{{Y{\i}ld{\i}z}(2014)}]{2014MNRAS.437..185Y}
{Y{\i}ld{\i}z}, M. 2014, \mnras, 437, 185, \dodoi{10.1093/mnras/stt1874}

\bibitem[{{Yildiz} \& {Do{\u{g}}an}(2013)}]{2013MNRAS.430.2029Y}
{Yildiz}, M., \& {Do{\u{g}}an}, T. 2013, \mnras, 430, 2029,
  \dodoi{10.1093/mnras/stt028}

\bibitem[{{Zhang} {et~al.}(2019){Zhang}, {Qian}, {Wu}, \&
  {Zhou}}]{2019ApJS..244...43Z}
{Zhang}, J., {Qian}, S.-B., {Wu}, Y., \& {Zhou}, X. 2019, \apjs, 244, 43,
  \dodoi{10.3847/1538-4365/ab442b}

\bibitem[{{Zhang} {et~al.}(2020){Zhang}, {Qian}, \&
  {Liao}}]{2020MNRAS.492.4112Z}
{Zhang}, X.-D., {Qian}, S.-B., \& {Liao}, W.-P. 2020, \mnras, 492, 4112,
  \dodoi{10.1093/mnras/staa079}

\bibitem[{{Zhou} {et~al.}(2016){Zhou}, {Qian}, {Zhang}, {Jiang}, {Zhang}, \&
  {Kreiner}}]{2016ApJ...817..133Z}
{Zhou}, X., {Qian}, S.~B., {Zhang}, J., {et~al.} 2016, \apj, 817, 133,
  \dodoi{10.3847/0004-637X/817/2/133}

\bibitem[{{Zhu} {et~al.}(2016){Zhu}, {Zhao}, \& {Zhou}}]{2016RAA....16...68Z}
{Zhu}, L.-Y., {Zhao}, E.-G., \& {Zhou}, X. 2016, Research in Astronomy and
  Astrophysics, 16, 68, \dodoi{10.1088/1674-4527/16/4/068}

\end{thebibliography}
\bibliographystyle{aasjournal}


\end{CJK*}
\end{document}